\documentclass[aps,pra,twocolumn,showpacs,superscriptaddress,amsmath]{revtex4}
\usepackage{longtable}
\usepackage{graphicx}
\usepackage{times}
\usepackage{amssymb}
\usepackage{amsmath}

\usepackage{color}

\newcommand{\be}{\begin{equation}}
\newcommand{\ee}{\end{equation}}
\newcommand{\beq}{\begin{equation}}
\newcommand{\eeq}{\end{equation}}
\newcommand{\bea}{\begin{eqnarray}}
\newcommand{\eea}{\end{eqnarray}}
\newcommand{\bdm}{\begin{displaymath}}
\newcommand{\edm}{\end{displaymath}}
\newcommand{\drm}{{\rm d}}

\begin{document}
\title{Ultra-high precision cosmology from gravitational waves}

\author{Curt Cutler}
\affiliation{Jet Propulsion Laboratory, California Institute of Technology, Pasadena, CA 91109, USA }

\author{Daniel E. Holz}
\affiliation{Theoretical Division, Los Alamos National Laboratory, Los Alamos, NM 87545, USA}

\date{\today}

\begin{abstract}
We show that the Big Bang Observer (BBO), a proposed space-based
gravitational-wave (GW) detector, would provide ultra-precise measurements of
cosmological parameters.  By detecting $\sim3 \times 10^5$ compact-star
binaries, and utilizing them as standard sirens, BBO would determine the Hubble
constant to $\sim 0.1\%$, and the dark energy parameters $w_0$ and $w_a$ to
$\sim 0.01$ and $\sim 0.1$, respectively. BBO's dark-energy figure-of-merit would
be approximately an order of magnitude better than all other proposed,
dedicated dark energy missions.

To date, BBO has been designed with the primary goal of searching for
gravitational waves from inflation, down to the level $\Omega_{GW} \sim
10^{-17}$; this requirement determines BBO's frequency band (deci-Hz) and its
sensitivity requirement (strain measured to $\sim10^{-24}$).  To observe
an inflationary GW background, BBO would first have to detect and subtract out
$\sim 3\times 10^5$ merging compact-star binaries, out to a redshift $z \sim 5$.
It is precisely this carefully measured ``foreground'' which would enable
high-precision cosmology.  BBO would determine the luminosity distance to each
binary to $\sim$ percent accuracy. In addition, BBO's angular resolution would be
sufficient to uniquely identify the host galaxy for the majority of binaries; a
coordinated optical/infrared observing campaign could obtain the redshifts.
Combining the GW-derived distances and the electromagnetically-derived redshifts
for such a large sample of objects, out to such high redshift, naturally leads
to extraordinarily tight constraints on cosmological parameters.  We emphasize
that such ``standard siren'' measurements of cosmology avoid many of the
systematic errors associated with other techniques: GWs offer a {\em
physics-based},
absolute measurement of distance.
In addition, we show that BBO would also serve as an exceptionally powerful
gravitational lensing mission,
and we briefly discuss other astronomical uses of BBO, including providing an early warning system for
all short/hard gamma-ray bursts.

\end{abstract}


\maketitle 

\section{Introduction}
\label{sec:intro}

Improving our understanding of the dark energy responsible for the observed
accelerating expansion of the Universe is one of the foremost challenges in
physics. Our current theoretical models are exceedingly inadequate. In addition,
given its extremely low energy density, there are no plausible scenarios for the
direct detection of dark energy. Thus progress on this critical issue must
be made through indirect observations.  In this paper we show that the Big Bang
Observer (BBO), a proposed space-based gravitational-wave (GW) mission designed
primarily to search for inflation-generated stochastic GWs in the band
$0.03\,{\rm Hz}$--$3\,$Hz~\cite{Phi2003}, would {\it also}\/ be an ultra-precise
cosmology mission, measuring the Hubble constant $H_0$ and the dark energy
parameters $w_0$ and $w_a$ far more accurately than other proposed dark energy
missions.

BBO has been proposed as a follow-on mission to the Laser Interferometer Space Antenna
(LISA), which will be most sensitive to GWs in the band $\sim
10^{-4.5}$--$10^{-1.5} {\rm Hz}$.  In the LISA band, any primordial GWs from
standard inflation are likely to be buried under the GW foreground from the
short-period white dwarf-white dwarf (WD-WD) binaries in the universe. The WD-WD
foreground is practically absent in the BBO band, since the WD-WD contribution falls very rapidly
for $f  > 0.01\,$Hz, and disappears entirely for $f > 0.25\,$Hz, where the most massive (and hence smallest)
WDs merge~\cite{vecchio_ungarelli,farmer_phinney,Phi2003}.
Instead, the dominant astrophysical foreground in the BBO band is mergers of compact binaries
composed of neutron stars (NSs) or black holes (BHs); i.e., NS-NS, NS-BH, and
BH-BH binaries. BBO would be sufficiently sensitive to individually
detect and subtract out essentially every merging compact binaries out to
high redshift, thereby uncovering any primordial GW background in its band that has energy density
$\Omega_{GW} \agt 10^{-17}$~\cite{Phi2003,CutlerHarms,Harms_etal_08}.

As far as the search for primordial GWs is concerned, merging compact binaries
represent a foreground that must first be removed. In this paper we show that
this foreground 
is a cosmological gold mine, allowing astronomers to
measure the expansion history of the Universe (out to at least $z \sim 5$)
far more accurately than the most ambitious currently-proposed cosmology
missions. The basic argument
is this: GW detectors in general, 
and LISA and BBO in particular,  will provide 
high-accuracy measurements of the luminosity distances, $D_L$, to detected merging binaries.
However, the GW signals contain (almost)
no information regarding the source
redshifts.  Thus the situation in GW astronomy is the exact reverse of
optical/electromagnetic astronomy: accurate distances will be relatively easy to
come by, while determining redshifts will
be much more challenging. 
For both the ground-based LIGO/VIRGO network and the space-based LISA, the angular resolution for
detected binaries will typically be several degrees~\cite{CuFl1994,Cutler98},
so the error box on the sky would contain $\sim 10^{5-6}$ possible host galaxies per event.
Therefore plans for doing
cosmology with GW observations have usually hinged on finding some electromagnetic outburst
associated with the  GW events, in order to identify the host galaxy and obtain its redshift~\cite{Schutz86}
(though see Finn and Chernoff~\cite{FinnChernoff93}, MacLeod and
Hogan~\cite{MacLeodHogan}, and Seto et al.~\cite{seto} for suggestions on how to
evade this requirement).  GW-detected merging binaries for which one can {\it
also} determine a redshift have been dubbed ``gold-plated'' binaries; Holz and
Hughes have shown that, with LISA, even a handful of such ``gold-plated''
detections of massive black hole binaries could make significant contributions
to cosmology~\cite{Holz_Hughes}.

BBO should detect $\sim 10^5$ merging NS-NS binaries per
year~\cite{CutlerHarms}, out to $z \sim 5$. BBO's angular resolution for NS-NS
mergers will typically be a few arcsec: a small enough error box to uniquely
identify the host galaxy in most cases.  This is especially true since one can
also use the source's luminosity distance, $D_L$ (measured to several percent),
and even a crude $D_L$--$z$ relation, to rule out galaxies at the right position
on the sky but with significantly different redshifts (e.g., differing by $>10\%$).  While
a unique host galaxy may not be
identifiable in very dense galaxy clusters, in such cases one can substitute the
average redshift of the cluster; this will lead to a typical redshift error of
order $\Delta z \sim 2\times 10^{-3}$, which is negligible for our purposes. 
Thus---in stark
contrast with LISA and ground-based GW detectors---BBO will detect $\sim 10^5$
``gold-plated'' binaries per year!  BBO will measure the luminosity distance
to each NS-NS binary with a relative accuracy of several percent.  For example,
for a NS-NS binary at $z=1.5$, the median distance error due to detector noise
will be $\sim 2\%$, while the distance error due to weak lensing (WL) will be
$\sim 7\%$.  For a large sample of sources, both of these errors ``average
out'', and the extent to which they do not average out can be readily modeled
and accounted for.  By contrast, the ambitious, space-based dark-energy mission
SNAP would be expected to observe roughly 2,000 SNe over the lifetime of the
mission, out to a maximum redshift of $z\sim1.7$~\cite{snap05}. 
BBO provides an overwhelmingly larger and deeper data set, with each individual distance
measured significantly more precisely. It is thus to be expected that BBO will
vastly outperform other proposed dark-energy missions; in fact, BBO might well
provide better constraints than all other proposed dark energy missions {\it combined}.
As we show below, with the inclusion of an (expected) Planck prior, BBO
data should measure 
$H_0$ to $\sim 0.1\%$, $w_0$ to $0.01$ and $w_a$ to $0.1$.

Perhaps even more importantly, we argue that the systematic errors associated
with GW detections are generally much smaller, and much easier to characterize,
than with any other proposed methods (e.g., weak lensing, supernovae, or baryon
acoustic oscillations). Type Ia supernovae (SNe) are arguably the state of the
art in cosmological distance measurement.  The intrinsic luminosity of a type Ia
SN can be {\em empirically}\/ calibrated to roughly 10\%. The physics underlying
this calibration is only poorly understood, and possible evolutionary systematic
effects are a grave cause for concern (see, e.g., \cite{sullivan, sdss,
two_pop}).  By contrast, the compact binaries detectable by BBO are
exceptionally simple sources.  At the orbital separations at which BBO observes
them, the compact objects can be treated as point masses (with spin), whose
dynamics are very accurately described by the post-Newtonian approximation.
Systematic distance errors arising from the detector itself will also be
negligible, since BBO (like LISA) is fundamentally
self-calibrating~\cite{science_case}.  The optical scheme is somewhat
complicated in practice, but in essence one is measuring the spacecraft arm
lengths (more precisely, small time-varying changes in the differences between
spacecraft arm lengths) in units of the laser wavelength. For LISA, the laser
wavelength (or equivalently its frequency) will be known to an accuracy of $\sim
10^{-6}\mbox{--}10^{-5}$; this small uncertainty is expected to dominate LISA's
calibration error, which is therefore also at the $\sim 10^{-6}\mbox{--}10^{-5}$
level.  (For the same reason, LISA Pathfinder's calibration is expected to be
accurate to better than $10^{-5}$ \cite{LTPnote}.)  One could reduce this small
inaccuracy by, say, stabilizing the laser to a transition line in an iodine gas
cell~\cite{Spero_private}, if there were sufficient motivation.  However there
is one caveat concerning BBO's calibration accuracy, which we discuss in
Sec.~\ref{caveat}.

In addition to determining the equation of state of dark energy, BBO should also
be an excellent probe of the growth of structure. BBO's measurements of
gravitational lensing will be as sensitive as dedicated weak lensing
missions. The basic idea is that once a $D_L$--$z$ relation has been derived
from the entire NS-NS sample, the dispersion of the $\sim 3\times 10^5$ NS-NS
data points about this curve is dominated by magnification by gravitational
lensing.  That is, one has obtained $\sim 3\times 10^5$ independent measurements
of the lensing magnification along different lines-of-sight, one for each
binary. We show below that the typical SNR for each measurement is $\sim 3.4$,
and therefore the SNR for the whole NS-NS population is $\sim 3.4\times
\sqrt{3\times10^5} \sim 2 \times 10^3$.  Although the NS-NS merger rate most likely exceeds the BH-BH rate, the lensing SNR
from the BH-BH population may exceed that of NS-NS mergers. The BH-BH merger rate is
poorly constrained by observation; a reasonable estimate from
population synthesis models is that the BH-BH rate is a factor $\sim 20$ smaller
than the NS-NS rate~\cite{sadowski}. However, the typical SNR for each BH-BH
merger is larger by a factor $\sim 5.3$ (for our fiducial NS and BH masses).
We thus estimate a lensing SNR for the whole BH-BH population of $\sim 2.2
\times 10^3$, and a total SNR for both populations combined of $\sim 3 \times
10^3$. This total lensing SNR could be significantly higher, if the BH-BH merger
rate is near the high end of the estimated range.  (Of course, advanced ground-based GW
detectors will {\it measure}\/ these rates, for $z \alt 0.4$, many years before
BBO flies.)  In Sec.~\ref{WL} we investigate BBO's
sensitivity as a gravitational lensing mission, and compare it with other
lensing missions.

While obtaining redshifts for $3 \times 10^5$ host galaxies would be a highly
ambitious goal at present, we are optimistic that it will be far less daunting by the
time BBO flies.  By then LSST may already have determined
photometric redshifts (accurate to $\sim 2$--$3\%$) for a large fraction of the
host galaxies in $\sim1/3$ of the sky.  In addition, there are many proposed wide-field
spectroscopic surveys; for example, BigBOSS would measure
$\sim 5$ million spectroscopic redshifts/yr (4,000 at a time) for galaxies in
the range $0.2 < z < 3.5$, over an area of $14,000\mbox{ deg}^2$~\cite{bigboss09}.
The success of BBO as a cosmological probe is dependent upon the determination
of host redshifts. It will thus be critical to secure the necessary optical
resources and develop an efficient strategy for obtaining redshifts for a
large sample of binary host galaxies.

After this work was mostly completed we found some brief remarks in the
literature that partly anticipate our results. Crowder \&
Cornish~\cite{CoCr2005} point out (in one sentence) that BBO should be able to
localize the host galaxy for most observed compact-binary mergers, but the
profound implications of this fact for physical cosmology are not developed.
One slide in a workshop talk on Decigo by Seto~\cite{seto-talk} lists in brief
bullets that Decigo measurements of NS-NS mergers could be used to probe dark
energy though the $D_L$--$z$ relation, and that some of these detections will
likely be GW-counterparts to gamma-ray bursts, allowing one to get a redshift.
But these slides give no estimates of the resulting cosmological constraints,
nor any sense of the potentially revolutionary implications of this observation
(perhaps because the current version of Decigo would likely have relatively poor
calibration accuracy, and so the cosmological measurements would suffer from
large systematics; see Sec.~\ref{caveat}). Seto et al.~\cite{seto} also
proposed using deci-Hz GW detections of inspiralling NS-NS binaries to measure
the universe's expansion, using a fundamentally different approach: observing
the small fractional change in source redshift over the course of the inspiral
(detected as a small time-varying phase shift). Because this phase shift is such
a small effect, the cosmological constraints that these authors estimated as
obtainable are orders of magnitude less sensitive than the constraints we find
(for comparable mission sensitivities).

In this paper we focus on BBO, although it is to be noted that the
Japanese GW community is proposing a similar mission called Decigo. 
Our basic conclusions apply to any deci-Hz GW mission
of roughly BBO-level sensitivity, including a Decigo-like mission, so long 
as the design ensures excellent calibration accuracy. We discuss this
further in Sec.~\ref{sec:decigo}.

The rest of this paper is organized as follows.  In Sec.~II we give a brief
overview of the BBO mission, its design sensitivity, the foreground produced by
NS-NS binaries, and how accurately BBO can measure the distance and sky location
of inspiralling compact binaries.  In Sec.~III we derive the cosmological
measurement accuracy obtainable by BBO from $\sim 3 \times 10^5$ NS-NS binaries.
We also investigate BBO's performance as a weak lensing mission, and discuss
some caveats accompanying our conclusions. In Section IV
we briefly discuss several other astrophysical uses of the BBO compact-binary
data: as an early warning system for {\it all} short/hard gamma-ray bursts, in
searches for the earliest intermediate-mass black hole mergers, and in studies of
hundreds of strongly lensed GW sources.  Our conclusions and plans for future
work are outlined in Sec.~V.

We use units in which $G=c=1$; everything can be measured in
the fundamental unit of seconds. However, for the sake of familiarity,
we also sometimes express quantities in terms of yr, Mpc, or $M_\odot$,
which are related to our fundamental unit by $1\mbox{ yr} = 3.1556 \times
10^7\mbox{ s}$,
$1\mbox{ Mpc} = 1.029 \times 10^{14}\mbox{ s}$, and $1 M_\odot = 4.926 \times
10^{-6}\mbox{ s}$.
For concreteness, in our simulations we adopt the following fiducial values
for the cosmological parameters: $H_0=70\,\rm km\,s^{-1}\,Mpc^{-1}$,
$\Omega_{\rm m}=0.3 $, and $\Omega_\Lambda=0.7$.

\section{Overview of BBO, Decigo,  and the NS-NS merger foreground}

\subsection{BBO}
BBO will be most sensitive in the band $\sim 0.03\mbox{--}3\,$Hz, which
is dictated by BBO's main design goal:
to detect primordial GWs generated by inflation.
BBO has been proposed as a follow-on mission to the Laser
Interferometer Space Antenna, which  will operate at $\sim
10^{-4.5}\mbox{--}10^{-1.5} {\rm Hz}$.
In the BBO band, the dominant astrophysical foreground will be mergers of NS-NS,
NS-BH, and BH-BH binaries.  The BBO mission is designed
so that essentially each and every merging compact binaries in the observable
universe (i.e., on BBO's past light cone) can be detected and subtracted out.
Cutler \& Harms~\cite{CutlerHarms} have presented a detection/subtraction
algorithm  and showed by an analytical calculation that it should
work; Harms et al.~\cite{Harms_etal_08} implemented the algorithm and demonstrated that it
worked well on simulated BBO data, albeit their demonstration was on a much
reduced data set because of computational limitations.  A very different
algorithm for detecting and subtracting the binary signals from BBO data is
also being developed by N. Kanda and collaborators (unpublished).

\begin{figure}[]
\centerline{\includegraphics[width=7.0cm]{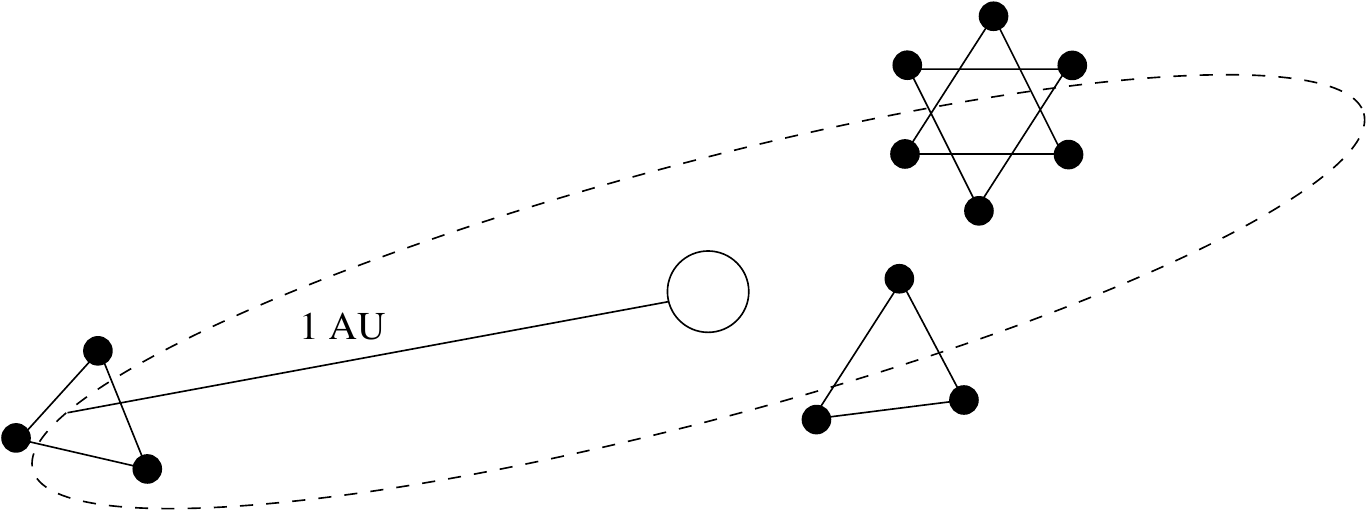}}
\vspace{0mm}
\caption[The BBO configuration]{Big-Bang Observer (BBO) consists
of four LISA-like triangular constellations orbiting the Sun at
$1\,$AU. The GW background is measured by cross-correlating the
outputs of the two overlapping constellations, while time-of-flight across the
Solar System gives BBO its angular resolution.   A schematic of Decigo (a
Japanese proposal similar to BBO) would be almost identical, except that the
constellations are $50$ times smaller than BBO's, and their arms form
Fabry-Perot cavities.
}
\label{figBBO}
\end{figure}

The current BBO design calls for four constellations of three satellites each,
all following heliocentric orbits at a distance of $1\mbox{ AU}$ from the Sun
(see Fig.~\ref{figBBO}).  Each 3-satellite constellation can be thought of as a
highly sensitive ``mini-LISA'', since the armlengths for each constellation are
two orders of magnitude smaller than LISA's, and BBO's sensitivity band is
correspondingly two orders of magnitude higher in frequency.  Two of the
constellations overlap to form a ``star of David''; the other two are ahead and
behind by $2\pi/3$ radians, respectively.  Briefly, the idea behind this orbital
geometry is that the energy density of the primordial GW background,
$\Omega_{\rm GW}(f)$, will be measured by cross-correlating the outputs of the
two overlapping constellations in the star of David (much as LIGO attempts to
measure $\Omega_{\rm GW}(f)$ by cross-correlating the outputs of the Livingston
and Hanford interferometers~\cite{AlRo1999}). The other two constellations give
BBO its angular resolution, which is useful for characterizing and removing the
merging compact binary foreground.  The source's angular position on the sky is
mostly determined by triangulation, using the differences in arrival times of
the GWs at the different constellations.

\begin{table}[b!]
\begin{center}
\begin{tabular}{|l|c|r|}
\hline
& Symbol & Value $\quad$\\
\hline\hline
Laser power & $P$ & $300\,$W \\
Mirror diameter & $D$ & $3.5\,$m \\
Optical efficiency & $\epsilon$ & $0.3$ \\
Arm length & $L$ & $5\cdot 10^7\,$m \\
Wavelength of laser light &$\lambda$ & $0.5\,\rm{\mu}m$ \\
Acceleration noise & $\sqrt{S_{\rm acc}}$ & $3\cdot 10^{-17}\,{\rm m}/({\rm s}^2 \sqrt{{\rm Hz}})$\\
\hline
\end{tabular}
\caption{BBO parameters} 
\label{tabParaBBO}
\end{center}
\end{table}

\begin{figure}[t!]
\hspace*{-0.5cm}\includegraphics[width=10.0cm]{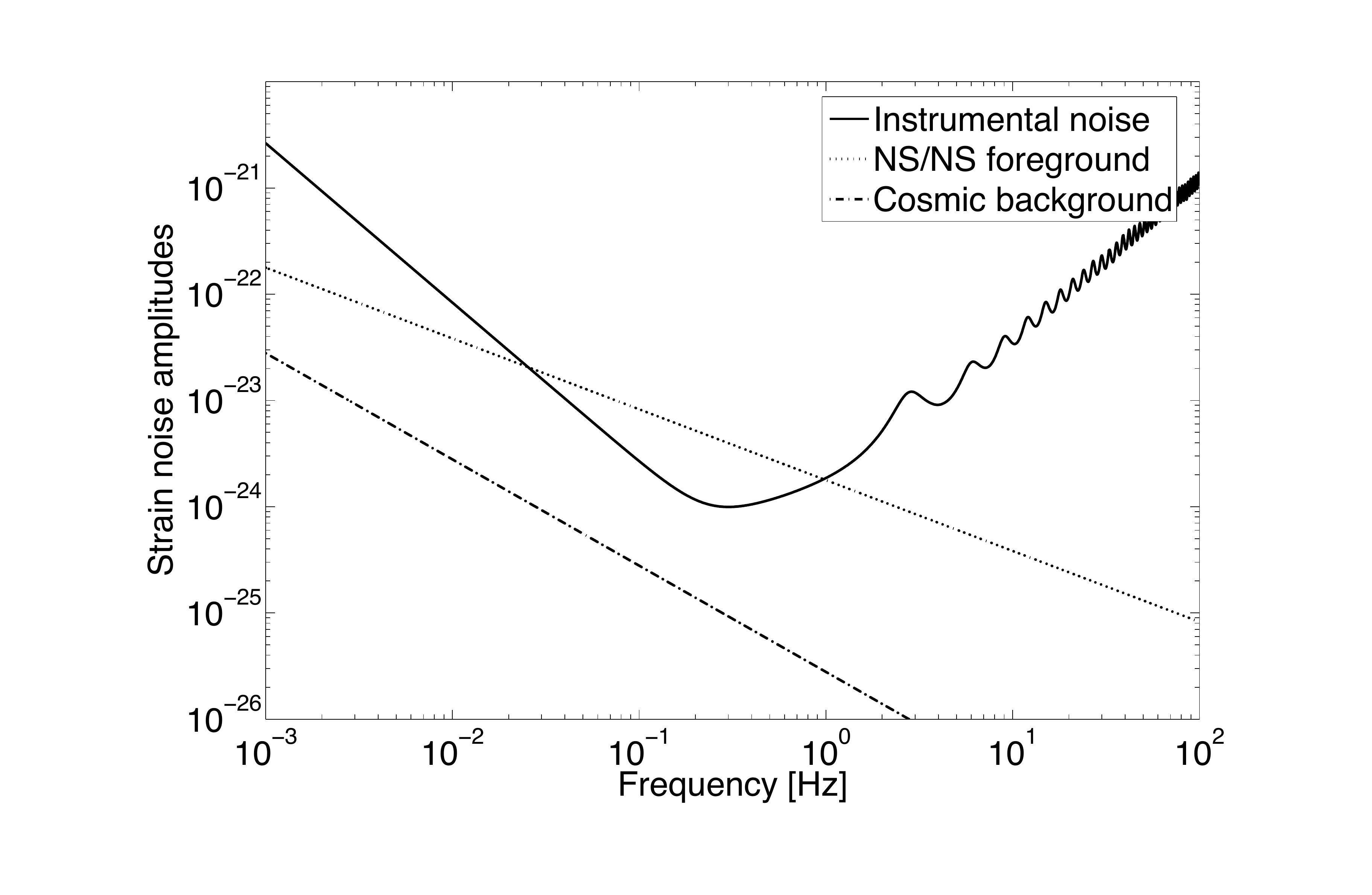}
\caption[BBO sensitivity curve]{Amplitude of BBO's
instrumental noise, $\sqrt{f S_{\rm h}(f)}$, compared to the amplitude of the
(pre-subtraction) NS binary foreground (plotted for $\dot n_0 = 10^{-7}\,{\rm
Mpc}^{-3}{\rm yr}^{-1}$) and the sought-for cosmic GW
background (plotted for $\Omega_{\rm GW}(f) = 10^{-16}$). To reveal this
cosmic background, the NS foreground must be subtracted off,
with fractional residual of $\lesssim 10^{-2.5}$. }
\label{figNoiseBBO}
\end{figure}

As explained in Cutler \& Harms~\cite{CutlerHarms} , for compact-binary mergers
the science output of each mini-LISA is in practice equivalent to the output of
two synthetic Michelson detectors, represented by the time-delay interferometry
(TDI) variables $X$ and $(Y-Z)/\sqrt{3}$.  We can therefore
regard BBO, which is made up of 4 mini-LISAs, as formally equivalent to 8
synthetic Michelson interferometers.  To construct the instrumental noise curve,
$S_{\rm h}(f)$, of {\it each} of these synthetic Michelsons, we use Larson's
on-line ``Sensitivity curve generator''~\cite{larson_online} and plug in BBO's
instrumental parameters, which are taken from the BBO Concept
Study~\cite{Phi2003} and also listed in Table~\ref{tabParaBBO}.  These
parameters will be subject to change as the mission evolves, but for
now they provide a convenient baseline.  The BBO Concept Study~\cite{Phi2003}
also lists parameters for less and more ambitious versions of the BBO mission,
referred to as ``BBO-lite'' and ``BBO-grand'', respectively, but in this paper
we restrict attention to the intermediate version, or ``standard BBO''.  In
using the on-line generator, we have specified that the high-frequency part of
$S_{\rm h}$ is 4 times larger than the contribution from photon shot noise
alone; this factor $4$ accounts for high-frequency noise components {\it other}
than shot noise, such as beam pointing jitter and stray light effects.
This is the
same choice made in Fig.~1 of the BBO proposal~\cite{Phi2003}, and is consistent with 
the standard assumptions made in drawing the LISA noise curve.
This BBO instrumental noise curve is shown in Fig.~\ref{figNoiseBBO}.

\subsection*{Decigo}
\label{sec:decigo}
BBO is seen as a follow-on mission to LISA in the U.S. and Europe, but in the Japanese GW
community there is a strong push to launch a deci-Hz GW mission first. The
current plan is for Decigo to be a factor $\sim 2$--$3$ less sensitive than BBO,  but for it
to launch earlier, in $\sim 2024$.  A small Decigo precursor mission, Decigo
Pathfinder, is among the final two missions competing for a launch slot in
$\sim$ 2012;  JAXA (the Japanese space agency) is scheduled
to decide in mid-2009 which mission gets this slot.

Our research to date has concentrated on BBO, but it would be straightforward to
generalize our BBO analysis to a mission with Decigo-level sensitivity. Indeed,
we expect that pursuing a BBO-style mission, but with a somewhat less ambitious
sensitivity goal, might be advisable from a cost/benefit standpoint; Decigo may
have a comparable cosmological reach to BBO, if designed to ensure excellent calibration accuracy.
In follow-up work, we plan to investigate how the science payoff from a deci-Hz
GW mission varies with its sensitivity.

\subsection{The NS-NS merger rate over time}
BBO would be able to observe BH-BH and BH-NS mergers to significantly higher
redshifts than NS-NS mergers, but the rates for BH-BH and BH-NS mergers are more
uncertain (and probably a factor $\sim 20$ lower) than NS-NS mergers,
so our discussion in this section will focus mostly on the NS-NS case.  The
extension of our work to NS-BH and BH-BH binaries is straightforward.

We denote the NS-NS merger rate (per unit proper time, per unit
co-moving volume) at redshift $z$ by $\dot n (z)$.
It is convenient to regard $\dot n(z)$ as the product of two factors:
\beq
\dot n(z) = \dot{n}_0\cdot r(z) \, ,
\eeq
where $\dot{n}_0$ is the merger rate today and $r(z)$
encapsulates the rate's time-evolution.
For $r(z)$, we adopt the following
piece-wise linear fit to the rate evolution estimated in \cite{SFMPZ2001}:
\beq\label{rz}
r(z) = \begin{cases}1+2z & z\leq 1\\
\frac{3}{4}(5-z) & 1\leq z\leq 5 \\ 0 & z\geq 5\end{cases}
\eeq
The current NS-NS merger rate, $\dot{n}_0$, is also usefully regarded as the
product of two factors: the current merger rate in the Milky Way, and a factor
that extrapolates from the Milky Way rate to the average rate in the
universe. The NS-NS merger rate in the Milky Way has been estimated by several
authors; it is still highly uncertain, but most estimates are in the range
$10^{-6}$--$10^{-4}\,{\rm yr}^{-1}$ \cite{BKB2002,KNST2001,VoTa2003}.  To
extrapolate to the rest of the universe, Kalogera et al.~\cite{KNST2001}
estimate that one should multiply the Milky Way rate by $1.1$--$1.6 \times
10^{-2}\cdot h_{70}\,{\rm Mpc}^{-3}$. This factor is obtained by extrapolating
from the B-band luminosity density of the universe, and it is only a little
larger than the extrapolation factor derived by Phinney in \cite{Phi1991}.  For
the estimates in this paper we adopt a rough geometric mean of these
rates: $\dot{n}_0 = 10^{-7} {\rm Mpc}^{-3}{\rm yr}^{-1}$.

\begin{figure}[]
\includegraphics[width=8.5cm]{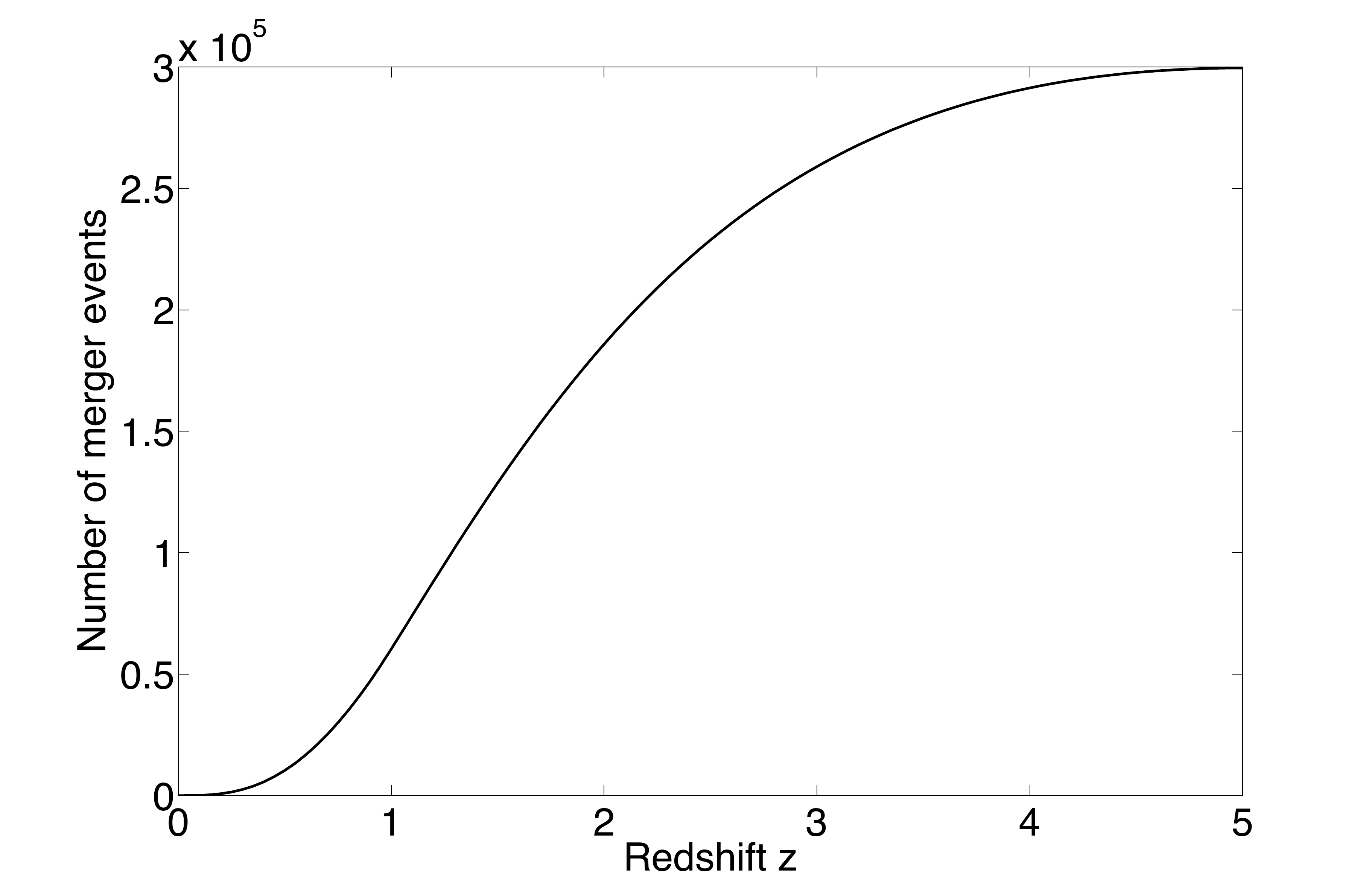}
\caption[]{The total number of NS-NS mergers closer than redshift $z$.
The results are normalized to a 3-yr observation period
and $\dot n_0= 10^{-7}\, {\rm Mpc}^{-3}{\rm yr}^{-1}$.}
\label{figNSNSnum}
\end{figure}

How many NS-NS merger events, $\Delta N_{\rm m}$,  enter the BBO band during some observation time,
$\Delta\tau_0$? Integrating the contributions from all redshifts, the rate 
is given by~\cite{CutlerHarms}:
\beq
\Delta N_{\rm m} = 3.0\cdot 10^5 \left(\frac{\Delta \tau_0}{3\,{\rm yr}}\right)\left(\frac{\dot n_0}{10^{-7}\,{\rm Mpc}^{-3}{\rm yr}^{-1}}\right).
\eeq
Based on our fiducial merger rate $\dot n(z)$, Fig.~\ref{figNSNSnum} plots the
number of observable mergers during three years of observation that occur closer than (any given)
redshift $z$. We see that roughly $15\% $ are at $z <1$, the median redshift is
$z \sim 1.6$, and roughly two-thirds are between $z =1$ and $3$.  This
distribution is well suited for probing the evolution of dark energy,
since it fully samples the evolution history during the transition from the matter
dominated to the dark energy dominated era~\cite{aldering06}.

The time required for a NS-NS inspiral signal to sweep through the BBO band
will typically be comparable to BBO's lifetime. More specifically,
the time remaining
until merger, from the moment the GW frequency sweeps through a given frequency,
$f$, is (to lowest post-Newtonian order)
\beq
t(f) = 4.64 \times
10^5\,{\rm s} \left(\frac{\mathcal{M}(1+z)}{1.22
M_{\odot}}\right)^{-5/3} \left(\frac{f}{1 {\rm Hz}}\right)^{-8/3}
\eeq
where $\mathcal{M}\equiv \mu^{3/5}M^{2/5}$ is the
so-called ``chirp mass'' of the binary (with $M$ the binary's
total mass and $\mu$ its reduced mass). For two $1.4
M_{\odot}$ NSs, $f \approx 0.205\,{\rm Hz},\,0.136\,{\rm
Hz},\,$ and $0.112\,{\rm Hz}$ at one year, three years, and five years before
merger, respectively.

\subsection{BBO's accuracy of parameter estimation}
Although it is straightforward to calculate BBO's angular resolution and distance measurement
accuracy using the Fisher matrix approximation, for pedagogical reasons we
begin with some simple, back-of-the-envelope estimates.
If the source's luminosity distance, $D_L$, were the
only unknown parameter, it could be measured to a relative accuracy of
$1/\mathrm{SNR}$, where SNR is the (amplitude) signal-to-noise ratio of the
detection.  For a merging NS-NS binary at $z=1$, the median SNR will be roughly
$180$, suggesting $\Delta D_L/D_L \sim 0.6\%$.  This is a lower limit on the
error, since correlations with other unknown parameters increase the
uncertainty.  Experience from similar problems in GW data analysis suggests an
increase by a factor of $\sim 2$, or $\sim 1\%$ distance uncertainty at $z=1$. 

The source's angular position on the sky will be inferred mostly by triangulation, based on
the GW time of flight between the different mini-LISA constellations.  This
idealization suggests the estimate $\Delta \theta \sim (1/\mathrm{SNR})
(1/2\pi\, 500\,f_0) \sim 1\, {\rm arcsec}$, where $f_0 \sim 0.3\,$Hz is BBO's
most sensitive frequency and we have used 1 AU $\approx 500\,$s in geometric units.  This estimate neglects both 
the information about the source's sky location that is encoded in the waveform by the
time-varying antenna patterns of the constellations, as well as the increased uncertainty 
that results from correlations with the other unknown parameters.

These back-of-the-envelope values turn out to be reasonably good
estimates of BBO's distance and angular position accuracies. We now turn
to a more careful, Fisher matrix calculation of BBO's parameter estimation accuracy.
The application of Fisher matrix techniques to LISA measurements of inspiralling
binaries was demonstrated in detail in Cutler~\cite{Cutler98}.  The
generalization to BBO, which is just four mini-LISAs with higher sensitivity, is
straightforward; our exposition will therefore be brief, and we refer the reader
to ~\cite{Cutler98} for more detail.  As mentioned previously, we can regard
BBO's output as formally equivalent to that of 8 independent, synthetic
Michelson interferometers; we represent BBO's output as
$s_{\alpha}(t)$, for $\alpha = 1 \cdots 8$.  We use ${\bf s}$ to
abstractly represent these $8$ time-series.  For simplicity we assume that the
detector noise is stationary, Gaussian, and the same for all four mini-LISA's
(in practice, we expect that the noise levels will be somewhat different and
slowly time-varying, in a manner that we can fit for).  Under these assumptions,
we obtain the following natural inner product on the vector space of signals.
Given two signals ${\bf g}$ and ${\bf k}$ we define $\langle {\bf g} \, | \,
{\bf k} \rangle$ by
\begin{equation}
\label{inner}
\langle {\bf g} \,|\, {\bf k} \rangle = 2
\sum_{\alpha=1}^{8}\int_{-\infty}^{\infty}\drm f\, \frac{(3/20) \tilde g_{\alpha}^*(f) \tilde k_{\alpha}(f)}{  S_{\rm h}(f) } \, .
\end{equation}
where $\tilde g_{\alpha}(f)$ and $\tilde k_{\alpha}(f)$ are the Fourier
transforms of $g_{\alpha}(t)$ and $k_{\alpha}(t)$, respectively.  We follow the
usual convention of taking $S_h(f)$ to be the {\it single-sided, sky-averaged}\/
noise spectrum for each synthetic Michelson interferometer.  The factor $3/20$ in
Eq.~(\ref{inner}) is the product of a factor $1/5$ due to the sky-average
convention and a factor $3/4 = {\rm sin}^2(\pi/3)$ arising from the $\pi/3$
angle between the arms in each constellation; see Sec.~V.A of
Barack \& Cutler~\cite{BC04} for a fuller explanation.

In this notation, the rms SNR for any waveform $\bf h$ is
\begin{equation}
\label{snh}
{\rm SNR}[ {\bf h}] = \langle {\bf h} \,|\, {\bf h} \rangle^{1/2} \, . 
\end{equation}
For a given incident gravitational wave, different realizations
of the noise will give rise to somewhat different best-fit
parameters.  However, for large SNR, the best-fit parameters will have a
Gaussian distribution centered on the correct values.
Specifically, let ${\tilde \lambda}^\mu$ be the ``true'' values of the
physical parameters,
and let ${\tilde \lambda}^\mu + \Delta \lambda^\mu$ be the best
fit parameters in the presence of some realization of the noise.  Then
for large $\rm SNR$, the parameter-estimation errors $\Delta \lambda^\mu$ have
a nearly Gaussian probability distribution whose covariance matrix
is given by
\begin{equation}
\label{bardx}
\overline{\Delta \lambda^\mu \Delta \lambda^\nu}
= (\Gamma^{-1})^{\mu\nu} \biggl(1+ {\cal O}({\rm SNR})^{-1}\biggr) \; ,
\end{equation}
\noindent where the overline ``$\overline{\ \ \ \ \ \ \  }$'' means ``expectation
value'', and where $\Gamma_{\mu\nu}$ is the Fisher matrix, defined 
by
\begin{equation}
\label{sig}
\Gamma_{\mu\nu} \equiv \left\langle \frac{\partial {\bf h}}{\partial \lambda^\mu}\, \Big| \,
\frac{\partial {\bf h}}{\partial \lambda^\nu }\right\rangle \; .
\end{equation}

Cutler and Harms~\cite{CutlerHarms} have shown that the effects of orbital
eccentricity on the NS-NS GW signal will typically amount to less than one
radian of phase over the entire $\sim 10^8$ radians of observed inspiral, while
the precession of the orbital plane due to the Lense-Thirring effect
will typically be $\alt 10^{-3}$ radians in the BBO band. We can therefore model
the binaries as quasi-circular, and neglect spin-precession effects (but we do
include spin-orbit effects on the waveform phase).  Our signal waveform,
$h_{\alpha}(t)$, thus depends on 10 physical parameters parameters describing
the binary: $\hat M_1$, $\hat M_2$, $\beta$, $\theta_S$, $\phi_S$, $\theta_L$,
$\phi_L$, $\phi_c$, $t_c$, and $ D_L$.  Here $\hat M_1$ and $\hat M_2$ are the
``redshifted mass'' ($\hat M_i = (1+z) M_i$), $\beta$ is a spin-orbit coupling
parameter defined in Cutler \& Flanagan~\cite{CuFl1994}, ($\theta_S$, $\phi_S$)
give the direction to the source, ($\theta_L$, $\phi_L$) describe the
orientation of the orbital plane, $D_L$ is the luminosity distance to the source, $t_c$ is the
time of merger, and $\phi_c$ is a constant of integration in the evolution of
the binary's orbital phase.
Our signal model is given by:
\begin{equation}
\tilde h_{\alpha}(f) = \frac{\sqrt{3}}{2} {\cal A} f^{-7/6} \Lambda_{\alpha}\bigl(t\bigr) e^{i\bigl(\Psi(f)- \varphi^p_{\alpha}(t) -  
\varphi^D_{\alpha}(t)\bigr)}  \; 
\ \ \ (f>0)
\label{h1}
\end{equation}
\noindent where
 $M \equiv M_1 + M_2$, $\mu \equiv M_1 M_2/M$, ${\cal M} 
\equiv \mu^{3/5} M^{2/5}$,  and
\be
{\cal A} \equiv  (5/96)^{1/2} \pi^{-2/3} D_L^{-1} \bigl[{\cal M} (1+z)\bigr]^{5/6} \, .
\ee
The relation between time and frequency, $t=t(f)$, is given through ${\cal
O}([v/c]^3)$ by
\cite{CuFl1994}
\FL
\begin{eqnarray}
\label{pntf}
t(f) & = & t_c - 5(8\pi f)^{-8/3} \bigl[{\cal M}(1+z)\bigr]^{-5/3}
\biggl[1 + \nonumber \\
\mbox{} & & {4\over 3}\left({{743}\over{336}} + 
{{11\mu}\over{4M}}\right)x - {{32\pi}\over 5} x^{3/2} +  O(x^2) \biggr] ,
\end{eqnarray}
and the waveform phase is given by
\begin{eqnarray}
&&\Psi(f)  = 
2\pi f t_c -\phi_c -\pi/4 +{3\over 4}\bigl(8 \pi {\cal M} (1+z) f
\bigr)^{-5/3} \nonumber \\
&& \times  \, \biggl[1+ {20\over 9}\left({743\over 336}+{{11
\mu}\over {4M}}\right)x +\left(4\beta -16 \pi \right)x^{3/2} \biggr] \;,
\end{eqnarray}
\noindent where the PN expansion parameter $x(f)$ 
is defined by 
\begin{equation}
\label{xf}
x(f) \equiv \biggl(\pi M (1+z) f \biggr)^{2/3} \; .  
\end{equation}
Equations for the modulation factor $\Lambda_{\alpha}(t(f))$ and ``polarization
phase'' $\varphi^p_{\alpha}(t(f))$  (both arising from the rotation of each
mini-LISA constellation as it orbits the Sun) as well as for the ``Doppler
phase'' $\Phi^D_{\alpha}$ (due to time delay between the passage of a particular
wavefront over the Solar System barycenter and its passage over each
constellation) are given explicitly in Secs. III and V of
Cutler~\cite{Cutler98}.  The exact expressions depend on the two angles
$(\beta_0,\alpha_0)$ that describe each constellation's position around the
Sun, and the orientation of each detector-triangle within its plane at some fiducial
time, $t_0$.  For definiteness, for the four mini-LISA's we choose: ($\beta_0,
\alpha_0)$ = $(0,0),
(0,\pi), (2\pi/3, 2\pi/3)$, and $(4\pi/3,4\pi/3)$.

The uncertainty in the source's angular position, $\Delta \Omega_S$
(in solid angle), is given by~\cite{BC04}
\begin{equation}\label{deloms}
\Delta \Omega_S  = 2 \pi\,\sqrt{(\Delta {\mu}_S)^2 \, (\Delta {\phi}_S)^2
- (\overline{ \Delta {\mu}_S \, \Delta {\phi}_S})^2 } \, .
\end{equation}
The $2\pi$ factor on the right-hand side (RHS) of Eq.~(\ref{deloms}) is
conventional; with this definition, the probability that the source lies {\it
outside} an (appropriately shaped) error ellipse enclosing solid angle $\Delta
\Omega$ is $e^{-\Delta \Omega/\Delta \Omega_S}$.  That is, as defined above,
$\Delta \Omega_S$ is very good approximation to the size of the $1\sigma$ error ellipse.

\begin{figure}[t!]
\centerline{\includegraphics[width=8.5cm]{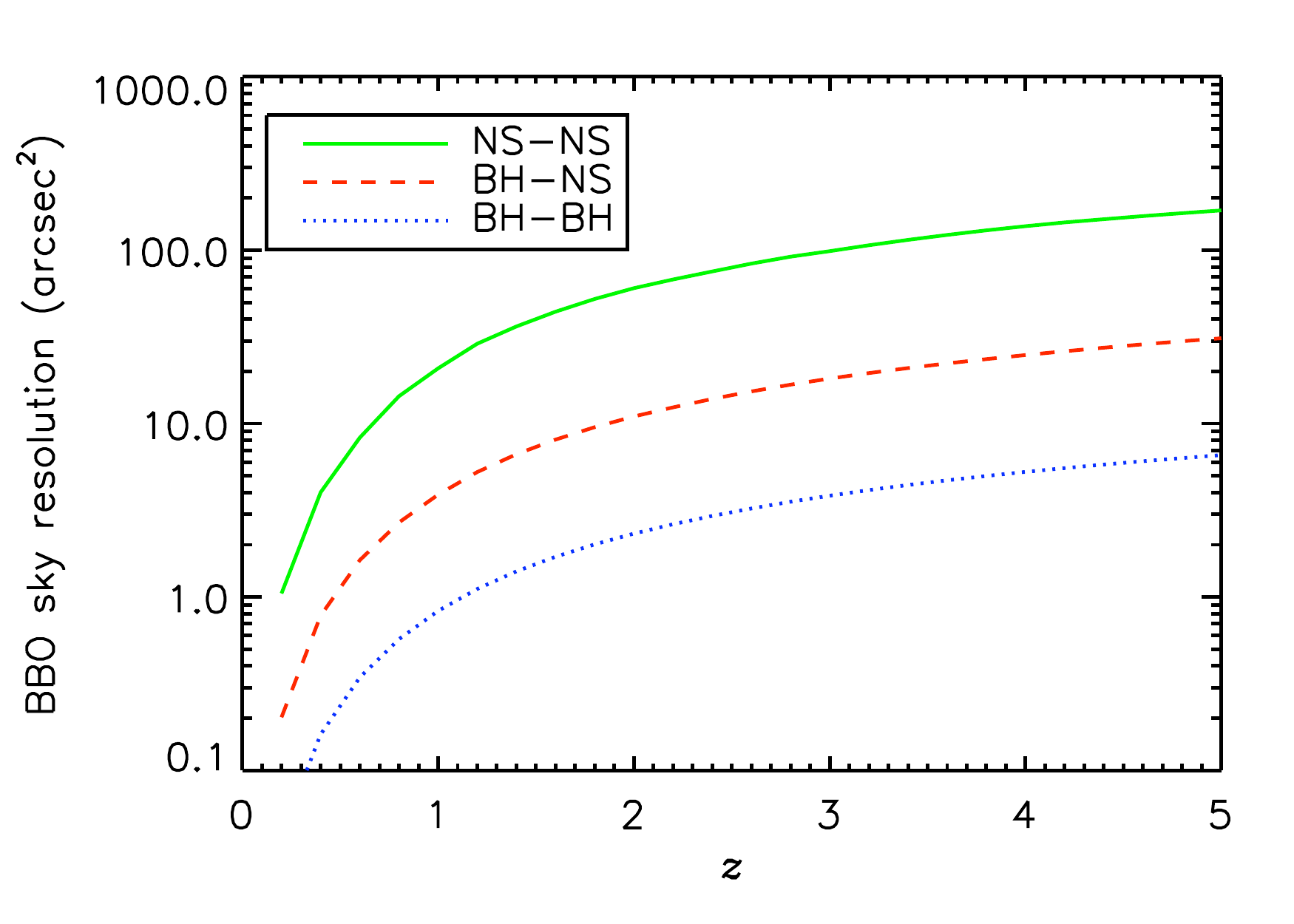}}
\caption[sky-accuracy plot]{BBO's angular resolution as a function of redshift,
$z$. The three curves show BBO's median $1\sigma$
angular resolution for three fiducial types of
merging compact binaries: BH-BH, BH-NS, and NS-NS. 
\label{figsky}
}
\end{figure}

\begin{figure}[b!]
\centerline{\includegraphics[width=8.5cm]{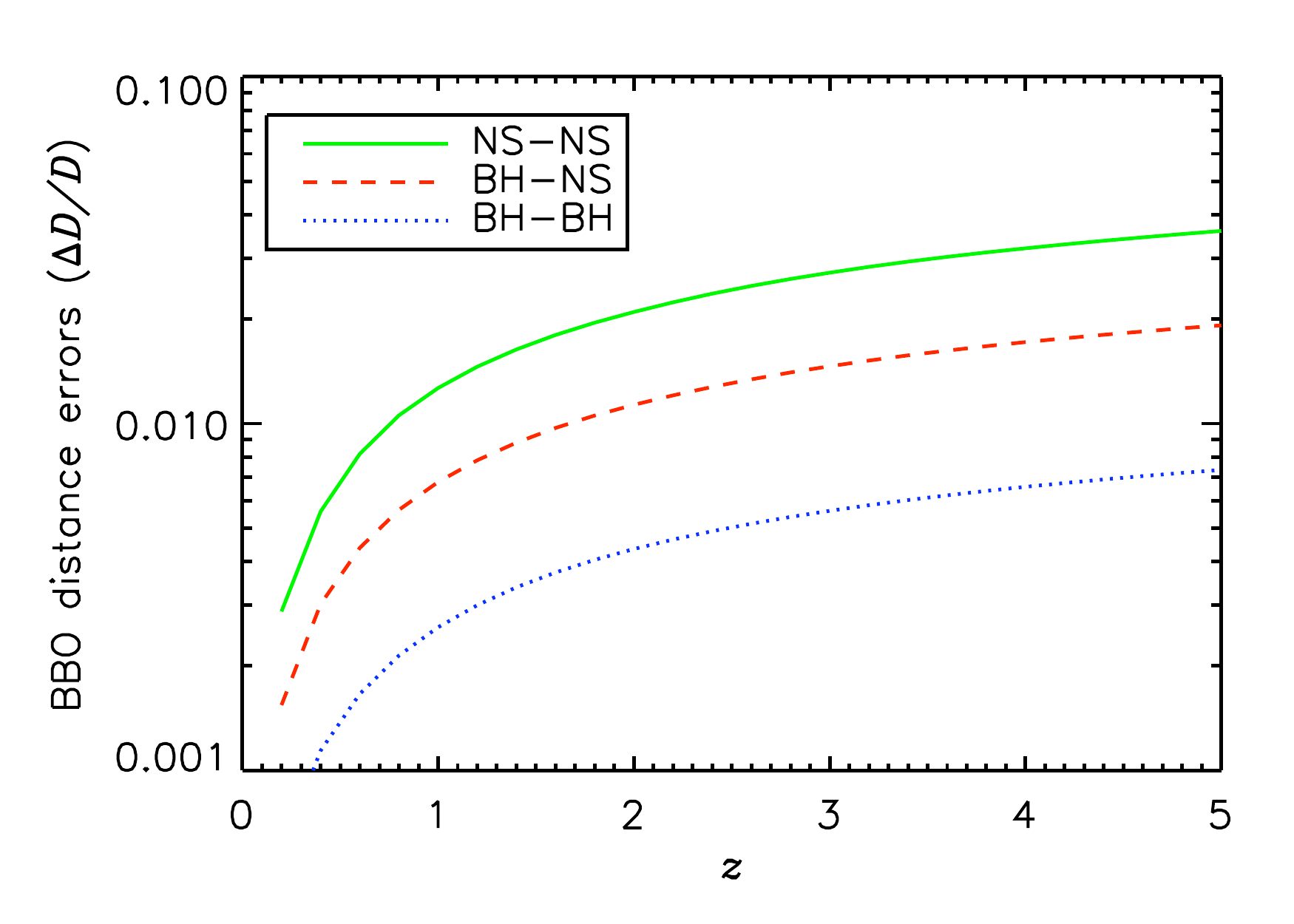}}
\caption[sky-accuracy plot]{BBO's median $1\sigma$ distance accuracy as a
function of redshift, $z$, for merging BH-BH, BH-NS, and NS-NS binaries.
} 
\label{figdist}
\end{figure}

BBO's median SNR angular resolution and distance accuracy (both $1\sigma$) for
NS and BH mergers and a range of $z$ are shown in Figs.~\ref{figsky} and
~\ref{figdist}, respectively.  These figures were produced as follows.  Each NS
was taken to have mass $1.4 M_\odot$, and each BH to have mass $10 M_\odot$.
For BBO's (sky-averaged) noise spectral density, $S_h(f)$, we adopted the fitting
function
\be\label{noise-fit}
S_h(f) = 6.15\times 10^{-51} f^{-4}  \ + \ 1.95\times 10^{-48}  \ + \ 1.2 \times 10^{-48} f^2  \, ,
\ee
\noindent where $f$ is in units of Hz.
For each $z$, we chose 250 random angle sets ($\theta_S$, $\phi_S$, $\theta_L$,
$\phi_L$), and computed the SNR, the Fisher matrix, and its inverse.  Since the
Fisher matrices are nearly degenerate, we tested robustness by using both
Matlab's standard matrix inversion function and Matlab's Cholesky-factorization
inversion routine; these were found to give essentially identical results.
While we argued above that the spin-spin coupling will have a negligible impact
on NS-NS waveforms for typical cases, as a further test of robustness we added
an additional spin-spin parameter (usually denoted ``$\sigma$'' in the
literature), and recomputed parameter estimation accuracies.  The results shown
in Figs.~\ref{figsky} and~\ref{figdist} turn out to be essentially independent
of the presence/absence of a spin-spin term in the waveform model.  Note that
the results in Figs.~\ref{figsky} and~\ref{figdist} are in reasonably good
agreement with the $z=3$ results in Table V of Cornish~\& Crowder~\cite{CoCr2005}
(considering that we model the high frequency part of BBO's noise curve somewhat
differently).

\begin{figure}[]
\includegraphics[width=8.5cm]{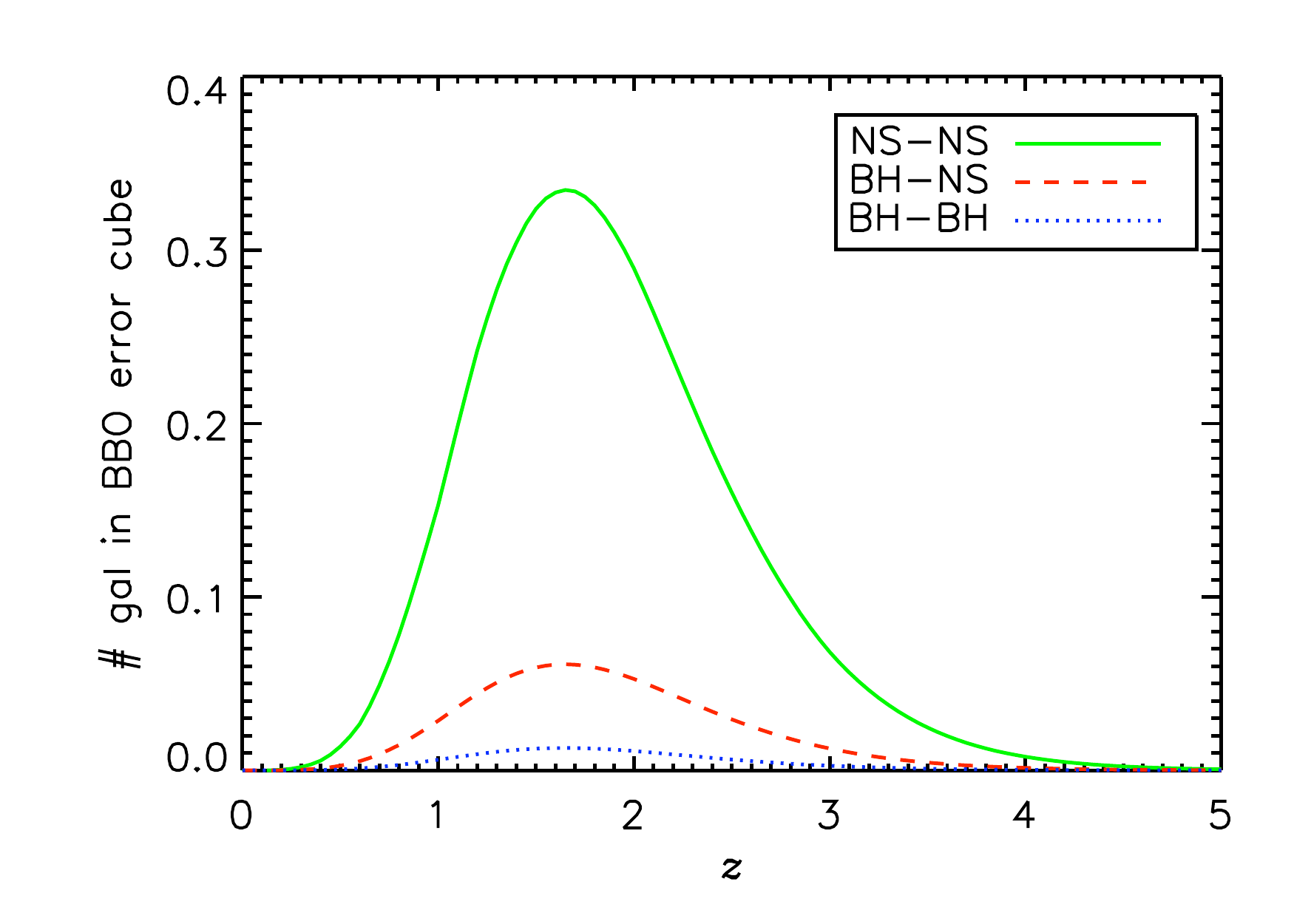}
\caption[]{Number of galaxies in the BBO error cube, as a function of
redshift. Even in the worst case, there is less than one galaxy within $1\sigma$
of a given binary on the sky, and therefore it should be possible to robustly
identify the unique host galaxy.}
\label{fig:galaxy_density2}
\end{figure}

Let us suppose that BBO identifies a binary system somewhere in the universe. We
now determine the number of potential host galaxies for the binary to be found
in the BBO error volume. We closely follow the approach of Holz \&
Hughes~\cite{Holz_Hughes}, updating their value for the projected number density
to the Hubble Ultra Deep Field number:
$dN/d\Omega= 1,000\mbox{ galaxies}/\mbox{arcmin}^2$~\cite{hudf}.
Since BBO measures distances at the percent level, the depth of the BBO error
box is dominated by the distance uncertainty due to gravitational lensing.
Following Eqs.~6--8 and Fig.~8 of~\cite{Holz_Hughes}, we calculate the total
number of galaxies in the BBO error box, per $\mbox{arcmin}^2$. We then multiply
this by the size of the BBO error box (shown in Fig.~\ref{figsky}), to arrive at
the total number of galaxies in the BBO error box, as a function of
redshift. This result is shown in Fig.~\ref{fig:galaxy_density2}.
The largest number of galaxies in a BBO error box is in the case of NS-NS
binaries at $z\sim1.5$, but even in this case there is less than ``half'' of a
galaxy present.
Thus even at the ``worst'' redshift, the median occupation fraction is less than
one---it will be possible to identify the {\em unique}\/ host for the majority
of BBO sources, and hence associate the appropriate redshift for the majority of
distance measurements. This is in contrast to the case of LIGO or LISA,
where the error boxes are large enough that associated electromagnetic activity
(such as a gamma-ray burst, or activity associated with a supermassive binary
black hole merger) is required to uniquely identify the
counterpart~\cite{Holz_Hughes,dalal06,nissanke09}. Obviating the need for an
independent identification of the counterpart sharply increases the expected
number of usable standard sirens, and hence significantly improves the accuracy
of the cosmological measurements.
Galaxy misidentifications will generally be seen as large outliers, and thus
their influence can be mitigated by the use of robust statistics, such as the Hough Transform. 

\section{Ultra-high precision cosmological parameters from BBO}

\subsection{The $D_L$--$z$ relation}
\begin{figure}[b]
\includegraphics[width=8.5cm]{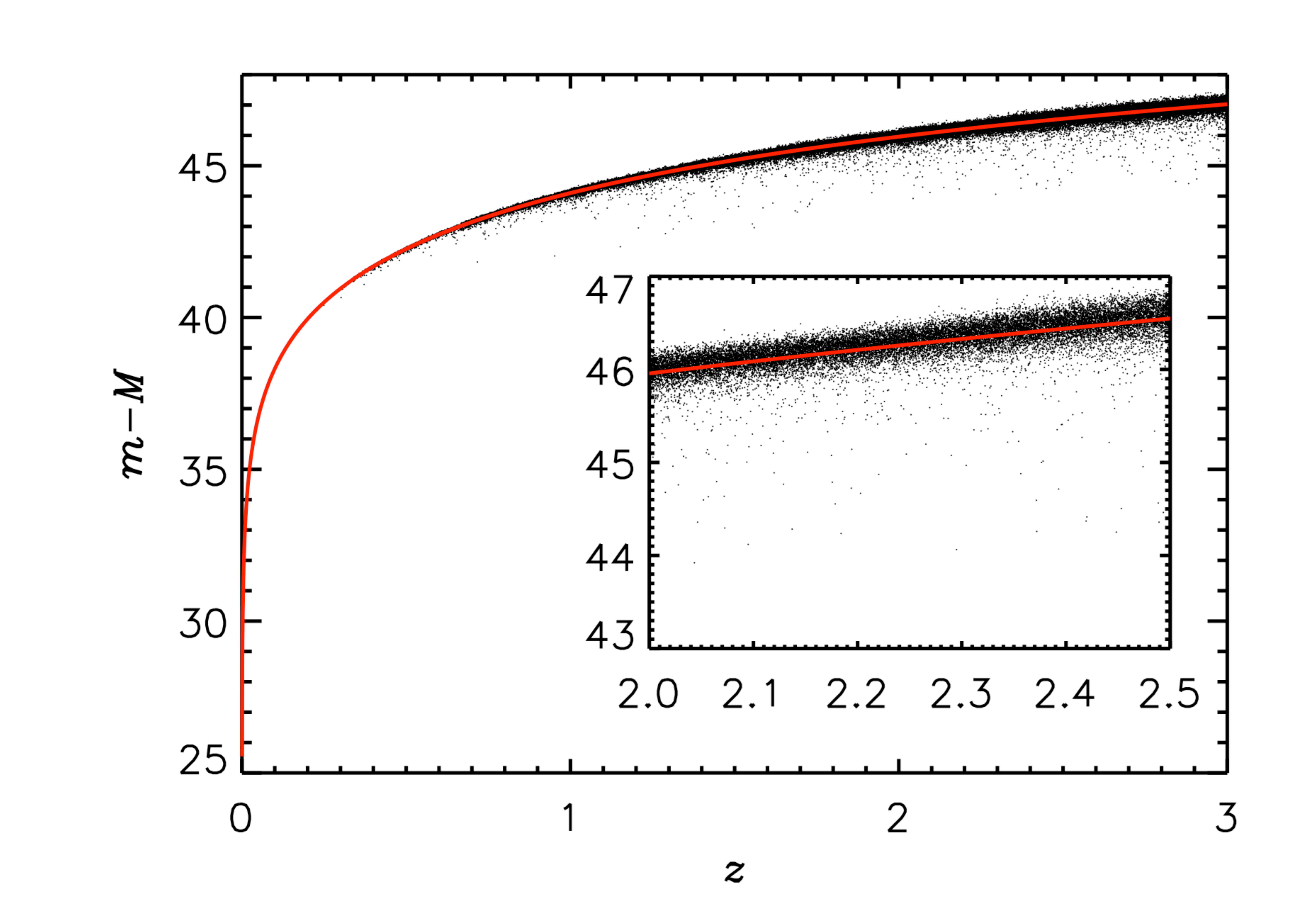}
\caption[]{Distance versus redshift for a sample BBO binary population. Distance
  is shown as distance modulus, and includes both BBO errors and gravitational
  lensing. The red curve is the true luminosity distance--redshift relation. Notice that
  lensing causes a small number of binaries to become tremendously magnified (to
  lower distance modulus), but there is a lower limit to the amount of
  de-magnification.}
\label{fig:scatter}
\end{figure}

We begin by considering BBO's measurements of the luminosity distance--redshift
relation (see Fig.~\ref{fig:scatter}). This relation is a direct measure of the evolution history of the
Universe: redshift provides the size of the Universe at emission, and luminosity
distance provides the time since emission. Thus a precise measurement of this
relation is sensitive to dark energy; indeed, it is this method that enabled the
initial discovery of the accelerating expansion of the Universe now associated
with dark energy.

We consider a fiducial population of $2.5\times10^5$ NS/NS binaries distributed
according to Eq.~\ref{rz}, out to $z=3$.  We assume that the distance
measurement errors due to detector noise for each individual binary are those
shown in Fig.~\ref{figsky}. Because BBO does such an exquisite measurement of
distance, the errors on the true distance to a given binary will be dominated by
the effects of gravitational lensing
magnification~\cite{Frieman:97,HolzWald:98}. We incorporate the lensing errors
following the approach of~\cite{holz_linder}, which is entirely appropriate
given the very high-number statistics we are considering. For each individual
binary we take the dispersion in flux due to lensing to be given by $\sigma_{\rm
lensing}=0.088z$ (see Eq.~9 of~\cite{holz_linder}). We have explicitly checked
that this approach is equivalent to drawing magnification values from the full,
non-Gaussian lensing probability distribution functions derived
in~\cite{HolzWald:98}.  We assume that the sky localization is sufficient for
the identification of a unique host galaxy (and hence redshift) for each binary
(as in Fig.\ \ref{figsky}).  The redshift determination will need to be done
independently of BBO, in the electromagnetic band.  While in practice there will
be some host galaxy misidentifications, for simplicity in this study we assume
that perfect redshifts have been obtained for all of our sources.  (This
simplification is partly based on our belief that a robust cosmological
parameter estimation method will substantially mitigate the effects of a
fractionally small set of misidentifications---enough so that in estimating
BBO's performance, to a first approximation it is reasonable to neglect them.)
We Monte Carlo generate populations of observed binaries, and then for each
population we determine the best-fit cosmological parameters (varying the number
of free parameters of interest). We repeat this procedure for a large ($>10^5$)
number of runs, and plot the resulting error contours. In what follows, the
$1\sigma$ contours contain $68.3\%$ of the best-fit values, and the $2\sigma$
contours contain $95.5\%$ of the models.

We follow the common convention of parameterizing the dark-energy equation of state in the two-parameter form~\cite{linder03}
\begin{equation}
w(z) = w_0 + w_a {z\over(1+z)}.
\end{equation}
We fit each data set to five cosmological parameters: the Hubble constant $H_0 =
h \times 100$ km/s/Mpc, the dark-matter density $\Omega_m$, the dark-energy
density $\Omega_x$, and the dark-energy phenomenological parameters $w_0$ and
$w_a$.  As is standard in assessing the power of proposed cosmology missions, we
include a forecasted Planck CMB prior, which constrains the angular diameter
distance at $z=1080$ to 0.01\%, and constrains $\Omega_mh^2$ to
1\%~\cite{verde07,mukherjee}.

\begin{figure}[t]
\includegraphics[width=8cm]{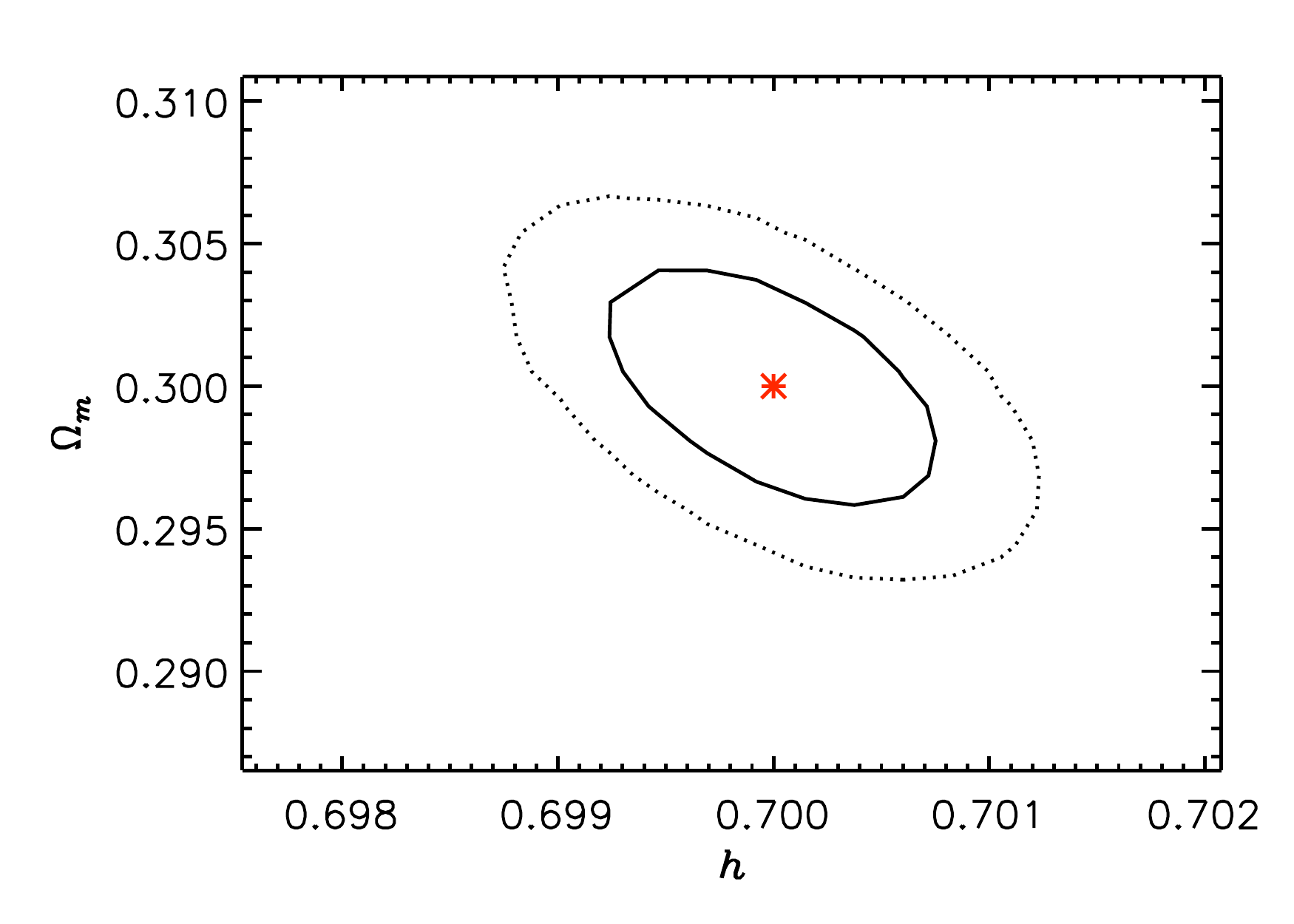}
\includegraphics[width=8cm]{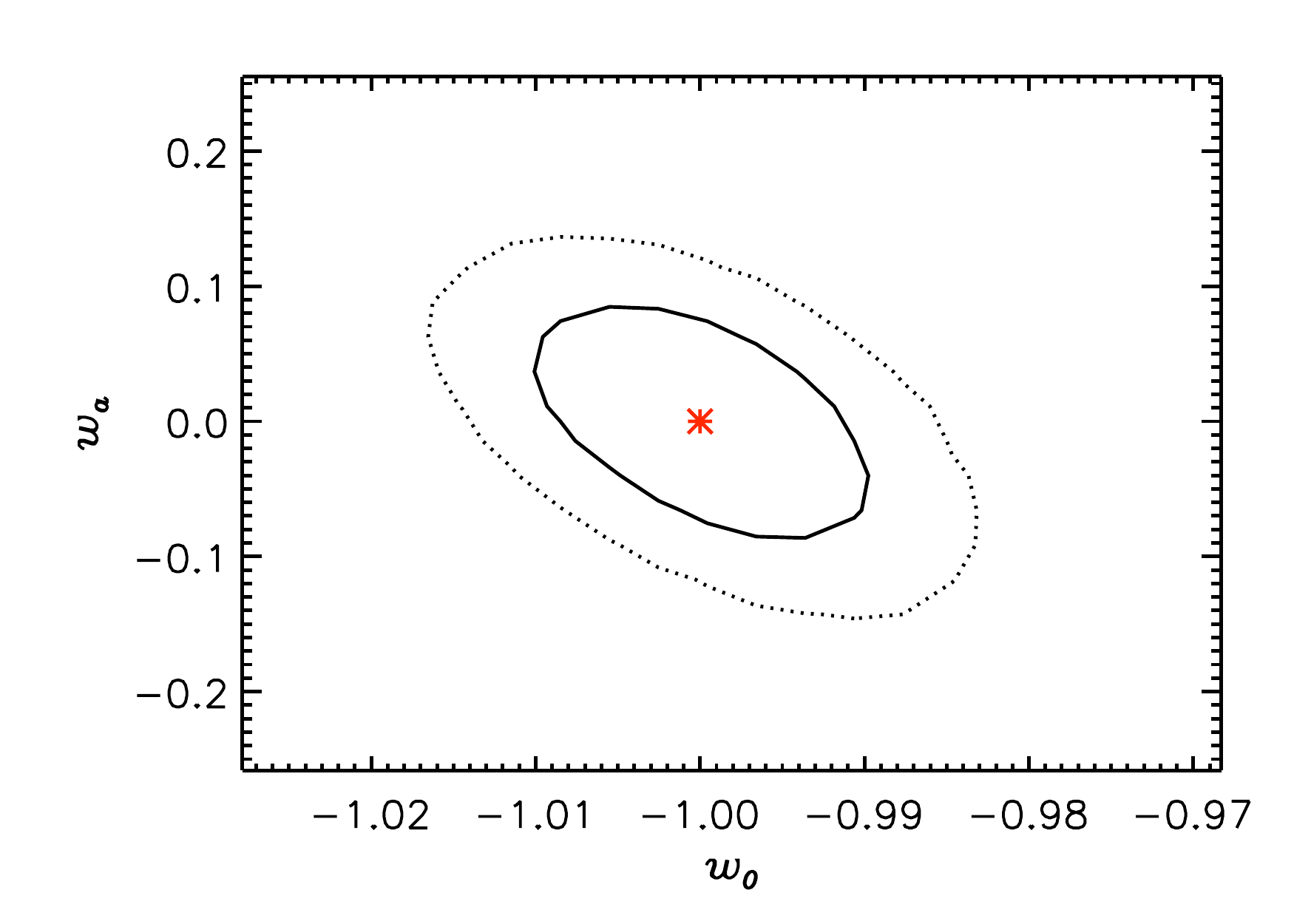}
\vspace{-0.5cm}
\caption[]{Top:  Measurement accuracy of the Hubble constant, $h$, and the
dark-matter density, $\Omega_m$. The solid and dashed curves map the $1\sigma$
and $2\sigma$ contours, respectively. The red star denotes the true underlying
model.  Bottom: Measurement accuracy of the
dark-energy equation-of-state parameters $w_0$ and $w_a$.\label{fig:w0_wa}}
\end{figure}

Fig.~\ref{fig:w0_wa} shows the resulting constraints on $h$ and $\Omega_m$,
assuming our fiducial population of binaries, and a 5-parameter fit to the data. We
find that BBO will measure the Hubble constant to $\sim0.1\%$, even when
marginalizing over two dark-energy parameters.  For comparison, the Hubble Key
Project (one of the major goals of the Hubble Space Telescope) arrived at a
value of $72\pm8\mbox{ km}/\mbox{sec}/\mbox{Mpc}$ ($>10\%$ error) for $H_0$
\cite{hst_hubble}. It is to be noted that, if we fit the data to a $\Lambda$CDM
model (e.g., setting $w_0=-1$ and $w_a=0$), we determine the Hubble constant to
$\sim0.025\%$. As recently emphasized in~\cite{riess09}, precision
measurements of the Hubble constant can be a key component of dark-energy
studies; BBO would provide the most precise measurement of $H_0$ that has ever
been contemplated.

In addition to the Hubble constant, BBO will directly constrain the dark-energy
equation of state. Figure \ref{fig:w0_wa} shows the BBO constraint on $w_0$ and
$w_a$, for our fiducial binary sample, with the inclusion of Planck CMB
priors. We find a $\sim 0.01$ constraint on $w_0$ and a $\sim 0.1$ constraint on
$w_a$. We note that we have not assumed a flat Universe in these fits, nor do we
incorporate any other cosmological measurements (beyond Planck). For comparison,
we consider the stage IV dark-energy missions (supernovae, baryon acoustic
oscillations, and weak lensing), as listed by the dark-energy task
force~\cite{detf}, representing the state of the art in future dark-energy
missions. The combination of
all stage IV missions improves the task-force figure of merit by a factor 8 to 15 with
respect to stage II missions (see pp.~18--20 and pp.~77--78 of \cite{detf}).
For comparison, BBO finds an equivalent figure of merit enhancement of
$\sim100$, roughly an order
of magnitude better than all of the stage IV missions, {\em combined}. It is
also to be emphasized that there are still fundamental concerns regarding
possible systematic errors in all of the stage IV missions, and thus their
combined figure of merit is undoubtedly optimistic. As discussed above, we
expect the systematic errors associated with BBO measurements to be negligible,
as it should be possible to build BBO such that calibration errors are much smaller than
$\sim 10^{-4}$.

\subsection{Weak Gravitational Lensing and Growth of Structure}\label{WL}
In addition to providing precision measurements of the fundamental cosmological
parameters ($H_0$, $\Omega_m$, $\Omega_k$, $w_0$, and $w_a$), BBO will also
directly measure the effects of gravitational lensing, and thus place
strong constraints on the primordial dark matter power spectrum, $P(k)$, and the
growth of structure. The growth of inhomogeneities is particularly sensitive to
gravity, and thus is a powerful way to constrain theories that modify gravity
as an alternative to assuming a dark-energy component.

One of the most powerful ways to measure the growth of density perturbations is
through gravitational lensing shear maps. This is done by observing the shapes
of large numbers ($\sim10^9$) of background galaxies, and measuring the subtle
correlations in the shapes of these galaxies due to the shear from gravitational
lensing.  The shear power spectrum at any redshift is sensitive not only to the
distances between observer, lens, and source (and thus, to the dark energy
component), but also to the distribution of lenses. This lens distribution is a
direct measure of the dark matter power spectrum as a function of redshift,
which is in turn sensitive to the growth function of perturbations, and thus the
gravitational force.

BBO would provide definitive measurements of the gravitational lensing
convergence power spectrum, comparable to state-of-the-art proposed measurements
of the lensing shear power spectrum. BBO measures an absolute luminosity
distance to each of the $\sim 10^5$ binaries. The error on this measurement is
almost entirely dominated by the effects of gravitational lensing
magnification. Once the average luminosity distance--redshift relation is
determined (as discussed in the previous section), it is possible to measure the
deviations from the background relation. Because the intrinsic uncertainty in
the distance measured by BBO is negligible when compared with lensing (see
Fig.~\ref{figdist}), each individual binary thus becomes a direct measure of
the gravitational lensing magnification along the given line of sight. The
population of binaries thus provides a few times $10^5$ individual measurements of the
magnification out to $z\sim3$. By evaluating the two-point correlation function
of these magnification measurements, it is possible to directly measure the
convergence power spectrum (which is equivalent to the shear power spectrum;
convergence, $\kappa$, is related to magnification, $\mu$, by $\mu = 1+2\kappa$
in the weak lensing limit). This approach has been discussed, for the case of Type Ia supernova
distance measurements, in~\cite{mag_map}. Here we follow an identical approach,
using binary standard sirens instead of supernova standard candles. In our case
each individual distance measurement is at least an order of magnitude
better, and we have an order of magnitude more sources, even compared to the
very ambitious supernova sample considered in~\cite{mag_map}. We note that in
what follows we focus on the weak lensing power spectrum, and for simplicity
neglect strong lensing. The latter will be discussed in more detail in
Section~\ref{sec:strong_lensing}.

In the Introduction we provided a rough estimate that BBO could measure weak
lensing (WL) with SNR of $\sim 2\times 10^3$ for its  NS-NS dataset and also
$\sim 2\times 10^3$ for its BH-BH dataset, for a total SNR of  $\sim 3\times 10^3$.
The JDEM design has not yet been determined, and the WL capability of the
mission varies quite significantly over the range of possibilities. The designs
that are best-suited for WL measurements contain $\sim 5$--$6 \times
10^8$ pixels in the focal plane and would have a goal of measuring galaxy
ellipticities to $\sim 0.1 \%$, and thus would require ellipticity correlation
measurements on $\sim 100$ galaxies to measure the WL effect to SNR of order 1.
(This is because galaxies typically have {\it intrinsic} ellipticites $\epsilon
\sim 0.3$, while the correlated ellipticity due to WL is a factor $\sim 10$
smaller, and SNR builds up as the square root of the number of galaxies
observed.)  Ideally, JDEM would measure shear for $\sim 10^9$ galaxies, covering
$\sim 10^4\mbox{ deg}^2$ on the sky, leading to a total SNR of $\rho_{\rm SNe}
\sim 3000$. We note that LSST is expected to measure weak lensing for
$\sim2\times10^9$ galaxies, out to $z=3$, over $\sim2\times10^4\mbox{ deg}^2$, and
is thus comparable to the most optimistic space-based lensing missions. These
estimates of the power of weak lensing shear measurements assume that
systematic errors (including telescope distortion, shear calibration,
point-spread-function correction, and redshift calibration) can be beaten down to
the $\sim 0.1\%$, which is quite optimistic (and far better than
is currently possible)~\cite{huterer06}.

The two methods of measuring WL are rather different---individual
magnification measurements versus correlated ellipticity measurements---and a
proper Fisher-matrix calculation is required to accurately compare the science
yield from either method. Such a calculation for BBO is now underway and will be
published in a follow-up paper.  But, crudely, we expect the ratio of
cosmological parameter estimation errors to be comparable to the ratio of SNRs
for the two methods, which is of order one, when BBO is compared to JDEM
missions designed to maximize the WL science.

We next calculate how accurately BBO could  measurement the convergence power spectrum.
We  following closely the approach of~\cite{mag_map}. The weak lensing angular power
spectrum for magnification can be written as
\begin{equation}
C_l^{\mu\mbox{-}\mu}=\int dr\, {W^2(r)\over d_A^2} P_{\rm dm}(k={l\over d_A},r),
\end{equation}
where
\begin{equation}
W(r) = 3\int dr'\, n(r')\Omega_m{H_0^2\over c^2 a(r)}{d_A(r)d_A(r'-r)\over
d_a(r')}.
\end{equation}
Here $r$ is the comoving distance, $d_A$ is the angular diameter distance,
$n(r)$ is the number density of binary systems (normalized so that $\int dr\,
n(r)=1$), and $P_{\rm dm}$ is the three-dimension dark matter power spectrum
(calculated following the approach of~\cite{PeacockDodds96}). The error on the
measurement of the magnification power spectrum is given by
\begin{equation}
\Delta C_l^{\mu\mbox{-}\mu}=\sqrt{2\over(2l+1)f_{\rm sky}\Delta
l}\left({C_l^{\mu\mbox{-}\mu}+{\sigma_\mu^2\over N_{\rm binaries}}}\right),
\end{equation}
where $f_{\rm sky}$ is the fraction of the sky covered by the survey, $\Delta l$
is the binning width in multipole space,
$\sigma_\mu$ is the RMS uncertainty of the magnification measurement from each
binary, and $N_{\rm binaries}$ is the surface density of the binaries. The first
term represents the error from cosmic variance, while the second term represents
the error from shot-noise.

\begin{figure}[t]
\includegraphics[width=8.5cm]{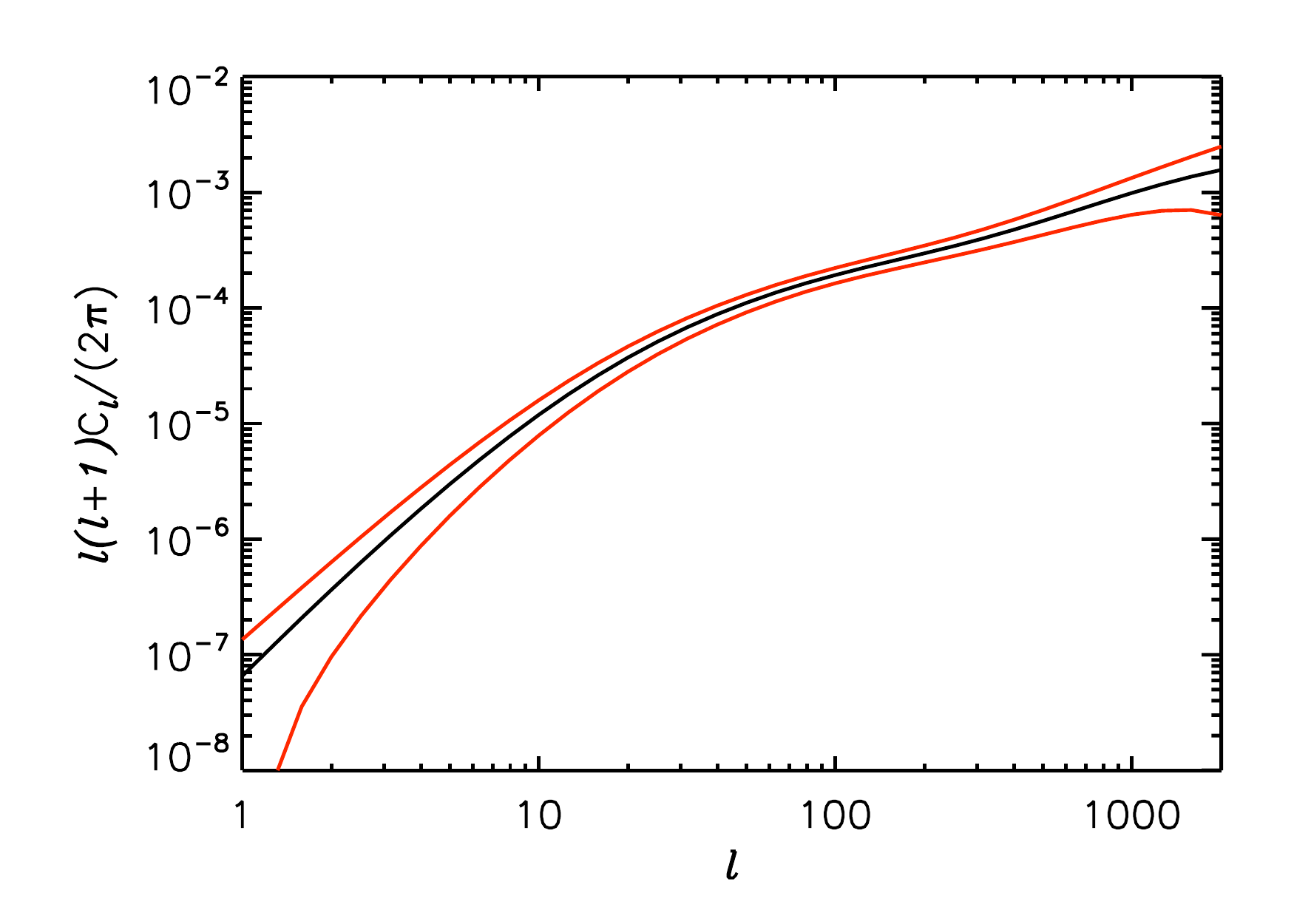}
\caption[]{BBO's measurement of the gravitational lensing convergence power
  spectrum, based just on NS-NS mergers. The red curves show the error bars, for each individual $l$
  mode. Each binary is a measure of the gravitational-lensing magnification
  (convolved with the intrinsic error) along that given line of sight. By
  observing many binaries, BBO produces a ``magnification map'' of the sky. This
  plot shows the power spectrum of this magnification map, which is sensitive to
  the growth of inhomogeneities in the Universe, as well as potential
  modifications to gravity.}
\label{fig:mag_map}
\end{figure}

Since the BH-BH merger rate is poorly known, to be conservative in the remainder of this
section we shall consider WL measurements of the NS-NS population only.
(Similar calculations for the BH-BH case, for a range of rates, will be published in later work.)
In Fig.~\ref{fig:mag_map} we show BBO's projected measurement of the
weak-lensing magnification map for NS-NS mergers.  We
note that the error bars are for each individual $l$ mode; no binning has been
performed in this figure. Defining the signal-to-noise ratio for this
measurement as
\begin{equation}
{S\over N} = \sqrt{\sum_l\left({C_l^i\over \Delta C_l^i}\right)^2}
\end{equation}
we find ${S/N}\approx120$, over an order of magnitude improvement over the
equivalent measurement using 10,000 SNe over 10 $\mbox{deg}^2$~\cite{mag_map}.

We emphasize that our technique for measuring WL through the magnification of GW sources is
entirely independent from the more traditional WL shear measurements, which
observe the correlations in galaxy shapes. Since the systematics are a major
source of concern for shear measurements, an equally powerful but
fundamentally independent technique for measuring WL would be highly
desirable. It will also be of interest to cross-correlate the two independent
measures of WL (amplification and shear). This will enable interesting
consistency tests, and directly test for the presence of systematics. In
addition, fundamental tests may be possible, since the
relation between shear and amplification is a robust, and thus far untested,
prediction of general relativity.

\subsection{Calibration issues and galaxy misidentifications}\label{caveat}

We have argued that BBO has revolutionary potential as both a dark energy and a
gravitational lensing  mission. There are a number of potentially important
systematics which must be addressed, foremost of which are ensuring calibration accuracy
and the potential misidentification of host galaxies (and hence redshift of the
binaries). We comment on these in turn.

The analysis in this paper is based on the basic BBO design put forth in the BBO
Concept Study~\cite{Phi2003}.  That design was extremely ``LISA-like'' in that
the test masses are freely floating, with no forces applied along the arm
directions.  In a later paper, Harry et al.~\cite{harry_etal} pointed out a flaw
in the Concept Study design---the laser power arriving at the photodiodes would
saturate them---and proposed a shift to a more ``LIGO-like'' design, with
forces applied to the test masses parallel to the arm axes, to keep the
photodiode operating near a dark fringe.  Because it is difficult to measure
this applied force accurately, this redesign would compromise the
self-calibrating quality that is one of LISA's strong points. LIGO's strain
calibration is accurate to within $\sim 8\%$, which is well below
the desired level for the standard siren measurements considered here. An author
of~\cite{harry_etal} has indicated that calibration
issues were not a major consideration during the re-design; until the work described in this
paper, the motivation for a highly accurate BBO calibration has not been
recognized \cite{Fritschel_private}.  We have consulted with interferometry
experts, and they suggested several plausible solutions to the saturation
problem that would preserve the LISA-like calibration
accuracy~\cite{Shaddock_Spero_private}.  One possibility is to use optics to
widen the beam, spreading the interfered light over an array of photodiodes.
Another possibility is to keep the latest LIGO-like design, but introduce an
additional small cavity behind the test mass to calibrate the force
applied~\cite{Shaddock_Spero_private}.  Our confidence that a good technical
solution can be found is enhanced by the fact that LISA-level calibration
accuracy is actually overkill for BBO; relative errors of $10^{-4}$, and perhaps
larger, are acceptable for the cosmological applications described here.

We also note that the current Decigo design is decidedly ``LIGO-like'' (or
``TAMA-like''), with far greater laser power in the arms than for BBO, and so
would have to be modified to achieve the calibration accuracy required for doing
ultra-high precision cosmology.  We hope that the prospect of precision
cosmology will spur instrumentalists to publish improved designs for
BBO and Decigo that have (or approach) LISA-level calibration accuracy.

Another possible source of error arises from the misidentifications of the galaxies
hosting the compact-binary mergers.  Misidentifications could arise for several
reasons: the field is particularly crowded; the position error is several
$\sigma$ (for $3\times 10^5$ sources, one expects position errors to range up to
$\sim 4.5\sigma$); or the host galaxy is very dim, and so the binary is
incorrectly ascribed to a brighter galaxy within the error box.  There are
strategies for mitigating some of these effects.  For instance,
misidentifications would naively seem to be most troublesome when they lead to
large outliers, but, for estimating cosmological parameters like $H_0$, $w_0$,
and $w_a$, one could clearly employ robust statistics that diminish the effect
of such outliers. Also, since for each detection one will {\it know}\/ the size
and shape of the error box and the density of galaxies within it, one should be
able to re-weight the data based upon the confidence of the identification.  This
re-weighting would enhance the importance of BH-BH mergers and gamma-ray
bursts, since in those cases the error volumes will be extremely small.  Of course,
misidentifications will lead to some degradation of BBO's WL results as well,
but the impact here is probably less, since a good fraction of the WL SNR will
likely come from BH-BH mergers, and BBO's angular resolution for these is $\sim
25$ times better (in solid angle) than for the NS-NS case.

The ``dim-galaxy'' problem is potentially more troublesome, since one is more
likely to miss a galaxy on the ``far side'' rather than the near side of the
error volume, which could lead to bias.  One could attempt to quantify this effect by comparing results for varying
exposure times (since in the limit of infinite exposure time, the dim-galaxy
problem goes away). We suspect that this bias is quite small, and it is to be
emphasized that this is a bias on the host identification (and hence redshift
determination), not on the primary observation of the GW binary. Addressing this
potential bias requires the development of a fairly detailed plan for searching
for optical counterparts, plus a detailed, robust data analysis algorithm that
mitigates misidentifications, which are beyond the scope of this paper.

\section{Further Astronomy from BBO}\label{other-astro}
Although BBO has been primarily conceived as a detector of inflation-generated GWs, in this paper
we argue that BBO would also be an unrivaled dark-energy mission.
In this section we briefly discuss some other unique
astronomical opportunities that would be afforded by BBO.

\subsection{Gamma-ray bursts}

Short/hard gamma-ray bursts are widely believed to result from (some subset of)
NS-NS or BH-NS mergers~\cite{belczynski_etal_06}.  If this is the case, then BBO
will serve as an ``early warning system'' for short/hard bursts, predicting the
precise time and sky location of {\em every}\/ burst, months in advance.  This
advance warning allows the bursts to be monitored with a full panoply of
telescopes {\em before}\/ they burst, and will permit searches for any
``pre-burst'' electromagnetic activity.  BBO will also tell us very precisely
the masses of the two bodies and the geometry of the system, which one will be
able to correlate with the electromagnetic signals.  This should fully resolve the question
of short/hard gamma-ray burst progenitors, as well as elucidate other important
issues (such as the nature and amount of beaming).  In fact, BBO may potentially
{\it overpredict}\/ the short/hard bursts, since not all mergers will
necessarily lead to observable bursts. Our knowledge of the orbital geometry of
the inspiraling binary should help predict which mergers will be observable
(e.g., if the gamma-ray beaming is perpendicular to the orbital plane).  BBO
would allow us to completely characterize the properties of orphan afterglows,
resulting from GRB events beamed such that they are not visible to us. In addition, BBO
would tell us which compact-binary mergers do {\it not}\/ lead to observable
bursts and/or afterglows.

It is to be noted that gamma-ray bursts will also serve as ``verification sources''
that confirm that BBO is working as expected: short/hard bursts should all ``go
off'' with merger times and sky-locations that are consistent with BBO's very
precise predictions.

\subsection{IMBHs to high redshift}

Another interesting (but more speculative) BBO source is the merger of
intermediate-mass black holes (IMBHs) at high redshift.  It is predicted that
the death of Population III stars could result in BHs weighing a few hundred
solar masses~\cite{heger_03}.  Some of these IMBHs may have grown (presumably
mostly by gas accretion) into the very massive BHs that now exist in the nuclei
of nearly all large galaxies. In the early build-up of galaxies from smaller
subhalos, when the subhalos merge the IMBHs they contain may be expected to merge as well.
BBO would detect these BH mergers with very high SNR, out to $z = 20$ and beyond. 
In Table~\ref{tabIMBH-snr2} we list BBO's SNR for IMBH mergers at $z=20$, for
several mass combinations.

\begin{table}[]
\begin{center}
\begin{tabular}{ | c ||  c | c |  c  | c | c | c |}
\hline
$M_1$ & $1\mbox{e}2$ & $3\mbox{e}2$ & $3\mbox{e}2$ & $1\mbox{e}3$ & $1\mbox{e}3$ & $1\mbox{e}3$ \\
\hline
$M_2$&$1\mbox{e}2$ &$1\mbox{e}2$ & $3\mbox{e}2$&$1\mbox{e}2$&$3\mbox{e}2$&$1\mbox{e}3$ \\
\hline
med SNR&$1.4\mbox{e}3$&$1.9\mbox{e}3$&$2.7\mbox{e}3$&$1.6\mbox{e}3$&$2.2\mbox{e}3$&$2.2\mbox{e}3$\\
\hline
\end{tabular}
\caption{Median matched-filtering SNRs for inspiralling intermediate-mass 
black hole binaries (IMBHs) at redshift $z
= 20$.  The masses are the locally measured ones (i.e., {\it not}\/ redshifted
masses), given in units of $M_{\odot} $.
}
\label{tabIMBH-snr2}
\end{center}
\end{table}

Note that for binaries with GW frequency at the last stable orbit satisfying
$f_{LSO} \agt 0.5\,$ mHz (i.e., lying rightwards of the minimum in the BBO noise
curve), one expects the SNR of the inspiral waveform to scale like ${\cal
M}^{5/6}$.  However, requiring $f_{LSO} \agt 0.5\,$ mHz requires $M(1+z) \alt 2.2 \times
10^3 M_{\odot}$, which is satisfied only by the first column in the table.  That
is, the tabulated SNRs are the results of two competing effects: as the masses
increase, so does the signal amplitude, but simultaneously less and less of the inspiral
waveform lies in the sensitive portion of the BBO band.

M. Volonteri has supplied us with the results of her merger-tree model, based on
two assumptions: 1)~today's massive and supermassive BHs began as initial
seed BHs of a few hundred $M_{\odot}$, and 2)~these seeds formed at $z \agt 20$, and
the seed BHs grew efficiently from accretion.  Using Volonteri's model, we
estimate that BBO's detection rate for IMBH mergers in the mass range $M_1 + M_2 <
1,000 M_{\odot}$ would be $\sim 30/\mbox{yr}$, of which  $\sim 25/\mbox{yr}$ would be at
 $z > 10$.   By comparison, Sesana et
al~\cite{Sesana_09} calculate that a network of three third-generation
ground-based interferometers would detect IMBH mergers at a rate of $\sim
2/\mbox{yr}$ (based on a very similar merger-tree model of Volonteri), and almost all
of these detected IMBHs would be at $z < 8$. 
Thus BBO offers a unique opportunity
to directly observer the very first seed black holes in the universe.

\subsection{Strong Lensing}
\label{sec:strong_lensing}
In addition to a direct measurement of the convergence power spectrum, BBO will
measure the full magnification probability distribution function as a function
of redshift, including resolving the high-magnification tail. These
high-magnification effects can be very sensitive to the dark-matter (and baryon)
profiles in galaxies~(see, e.g.,~\cite{oguri09} for an optical version of this).
For example, galaxies with more concentrated, cuspy profiles may be more likely
to engender strong lensing than those with more flattened cores.

The fraction of multiply-imaged quasars is $\sim 10^{-3}$.  One expects a
comparable fraction of multiply-lensed NS-NS binaries: i.e., $\sim 300$
multiply-imaged binaries.  In the BBO dataset, these will appear as pairs of
binaries having nearly identical sky locations and essentially identical masses,
spins, and orientations (i.e., identical to within the error bars), but with
different apparent distances (according to the magnification of each image), and
with arrival times differences of order months to years. 
The observed rate of multiple-imaging is an important probe of dark-matter
density profiles, as well as the overall dark matter distribution (e.g.,
$\sigma_8$), and BBO will provide an extremely clean measurement.
BBO's determination of the
time-delays between the multiple images should be accurate to $\sim\mbox{ 0.1
sec}$, a fractional accuracy of better than one part in $\sim10^8$.  Thus independent
estimates of the Hubble constant from time delays will be possible for $\sim
300$ multiply-imaged mergers, modulo the standard difficulty of accurately
modeling the density profiles of the lensing galaxies.  In addition, the relative
magnifications should help constrain the dark matter density profiles, as well
as directly break the mass-sheet degeneracy.

\section{Summary, conclusions and future work}

Studies of the dark energy generally take one of two approaches: The first
approach is to measure the
luminosity distance--redshift relation (or angular diameter distance--redshift)
to high accuracy. Type Ia supernovae and baryon-acoustic-oscillation
measurements fall into this category. The second approach to exploring the dark
energy is to measure the weak-lensing shear power
spectrum, and infer values for the growth of structure and cosmological distance
ratios. In this paper we have shown that BBO will provide unprecedented
measurements of both the luminosity distance--redshift curve as well as the weak
lensing power spectrum.

As discussed in Sec.~\ref{sec:intro}, the success of BBO as a cosmological
probe is dependent upon obtaining redshifts for a large number of the host
galaxies to the binary sources. It is to be emphasized that any subsample of
galaxies is sufficient (e.g., LRGs), since there are no standard siren systematics expected to
be associated with the nature of the host galaxy. 
Although obtaining redshifts for $3 \times 10^5$ host galaxies is certainly a demanding
requirement, as mentioned in Sec.~\ref{sec:intro}, we expect that surveys such
as LSST or BigBoss may provide the required dataset in due course. 
Future work will be required to refine the estimates of BBO's performance given
here, and to consider ways that the mission design might be modified to improve
its price/performance ratio from a cosmological perspective. We hope that this
work will encourage GW instrumentalists to develop workable designs with
calibration accuracy comparable to LISA's.  We plan on improving the analyses
presented here in several ways.  First, in this paper we made simplifying
approximations in calculating BBO's parameter estimation accuracy.  Although we
expect that these simplifications will have little impact on the results, we
will explicitly check this by 1)~including the effects of non-zero
eccentricity and Lense-Thirring precession in our waveform model, 2)~going
beyond the Fisher matrix approximation in calculating the expected sizes of
the errors~\cite{vallis08},
and 3)~calculating  how cross-correlations between different binaries affects parameter estimation for each.
Regarding the cosmological constraints derivable from BBO's dataset, we intend
to 1) explore BBO's performance as measured by other dark-energy figures of
merit, 2) improve the sophistication of our method for incorporating priors
(and, in particular Planck CMB priors) into the BBO analysis, 3) incorporate
robust parameter estimation
methods that mitigate the effects of host galaxy misidentifications, and
4) generalize our approach to a model-independent, multi-parameter description
of the dark energy~\cite{HutererCooray05,Li07,Sarkar08}, instead of the
arbitrary two-parameter form ($w_0$ and $w_a$) considered here.
In addition, we will investigate whether there are
special synergies in cross-correlating BBO's gravitational lensing magnification measurements with
the weak-lensing shear measurements made by a JDEM-like mission.  We also plan to
further explore BBO's potential impact upon other areas of astronomy (some of which were
sketched out in Sec.~\ref{other-astro}).  Lastly, we will investigate how
various figures of merit vary with BBO's sensitivity and other mission design
parameters.  There are simple variants to the current design which may
retain many of the high-precision cosmology applications while considerably
reducing the price or risk of the mission.
For instance, having four constellations of mini-LISAs is overkill for
high-precision cosmology; three constellations are certainly sufficient, and two
may be sufficient, if combined with measurements from planned, third-generation
ground-based GW detectors (such as the Einstein Telescope~\cite{ET}).  In the
latter case, the mission would require six satellites instead of twelve.  And
while a noise curve at the level shown in Fig.~\ref{figNoiseBBO} is probably
necessary for detecting a stochastic GW background, it is somewhat
overqualified for the task of doing ultra-high precision cosmology: as currently
conceived BBO
could detect all NS-NS mergers out to $z=5$, while for doing high-precision cosmology
it is probably sufficient to detect and localize a reasonable fraction of
mergers out to $z \sim 2$.  Since BBO would surely be a multi-billion dollar
mission, it is important to look for ways of
significantly reducing mission cost while retaining a large fraction of the science.

In this paper we have shown that BBO is a particularly powerful mission, as it
will provide revolutionary measurements of both the luminosity
distance--redshift relation and the growth of structure (through gravitational
lensing measurements).  Unfortunately,
a deci-Hz GW mission like BBO
is, optimistically, at least fifteen years from being built.  Nevertheless, the power
of a BBO-like mission for precision cosmology---stemming from very precise,
unbiased distance measurements of $\sim 3\times 10^5$ NS-NS binaries to $z \sim
5$---is so revolutionary that BBO could represent the future
of high-precision cosmology.

\acknowledgements
CC's work was carried out at the Jet Propulsion Laboratory, California Institute
of Technology, under contract to the National Aeronautics and Space
Administration.  He acknowledges support from a JPL Research and Technology
Development grant, as well as support from NASA grant 06-BEFS06-31.  We thank
Peter Bender, Peter Fritschel, Salman Habib, Jan Harms, Noboyuki Kanda, Seiji
Kawamura, Lexi Moustakis, Hans-Reiner Schulte, Tom Prince, Jason Rhodes, Naoki
Seto, Daniel Shaddock, and Michele Vallisneri for helpful discussions.  We
particularly wish to thank Marta Volonteri for sharing her merger-tree results
(which we used to estimate BBO's detection rate for high-$z$ IMBH inspirals),
Dragan Huterer and Licia Verde for help with the inclusion of Planck priors, and
Lee Lindblom for his observation that BBO would be an all-sky monitor for short
gamma-ray bursts.


\end{document}